\documentclass[12pt]{article}
\usepackage{graphics}
\usepackage{graphicx}
\usepackage{amsfonts}
\usepackage{amsmath}
\usepackage[numbers,sort&compress]{natbib}
\begin{document}
\date{}

\title{Two types of generalized integrable decompositions and new solitary-wave
solutions for the modified Kadomtsev- Petviashvili equation with
symbolic computation}

\author{ Tao Xu$^{1}$,  Hai-Qiang Zhang$^{1}$, Ya-Xing Zhang$^{1}$, \\
Juan Li$^{1}$ and Bo Tian$^{1,\,2,\,}$\thanks{Corresponding author,
with e-mail address:
 gaoyt@public.bta.net.cn}
\\
\\{\em 1.  School of Science, P.\ O.\ Box 122, Beijing University of} \\
{\em  Posts and Telecommunications, Beijing 100876, China
} \\
{\em 2. Key Laboratory of Optical Communication and Lightwave} \\
{\em Technologies, Ministry of Education, Beijing University of} \\
{\em Posts and Telecommunications, Beijing 100876, China}
 } \maketitle

\begin{abstract}

The modified Kadomtsev-Petviashvili (mKP) equation is shown in
this paper to be decomposable into the first two soliton equations
of the $ 2N $-coupled Chen-Lee-Liu  and Kaup-Newell hierarchies by
respectively nonlinearizing two sets of symmetry Lax pairs. In
these two cases, the decomposed (1+1)-dimensional nonlinear
systems both have a couple of different Lax representations, which
means that there are two linear systems associated with the mKP
equation under the same constraint between the potential and
eigenfunctions. For each Lax representation of the decomposed
(1+1)-dimensional nonlinear systems, the corresponding Darboux
transformation is further constructed such that a series of
explicit solutions of the mKP equation can be recursively
generated with the assistance of symbolic computation. In
illustration, four new families of solitary-wave solutions are
presented and the relevant stability is analyzed.

$\;$

\noindent{\textit{Keywords}: Modified Kadomtsev-Petviashvili
equation; Integrable decompositions; Darboux transformations;
Solitary-wave solutions; Symbolic computation}

\end{abstract}

\newpage

\noindent {\textbf{1. Introduction}}

According to Lax's theory~\cite{a01}, a given NLEE is said to be
integrable
if it arises as the compatibility condition of two linear
eigenvalue equations which are usually called a Lax pair and
comprised of the spatial part and the temporal part. Although it
is not an easy work to find the Lax pair associated with an
integrable NLEE, one can relate a properly-chosen spectral problem
to a hierarchy of soliton equations whose Lax pairs have the same
spatial part but different temporal parts~\cite{a02}. In the past
several decades, some important and typical (1+1)-dimensional
integrable hierarchies have been established and well understood,
including the Ablowitz-Kaup-Newell-Segur (AKNS)~\cite{a03},
Wadati-Konno-Ichikawa (WKI)~\cite{a04}, Kaup-Newell
(KN)~\cite{a05} and Levi~\cite{a06} hierarchies. Today, the Lax
pair has been playing a considerable role in studying the
integrable properties of NLEEs such as the Hamiltonian structures,
conservation laws and symmetry classes~\cite{a07}.

For describing various complex nonlinear phenomena of our
realistic world, the higher-dimensional NLEEs appear very
attractive in many fields of physical and engineering
sciences~\cite{a07a}. However, due to the higher space dimensions,
those higher-dimensional nonlinear systems often exhibit more
intricate properties (e.g., the integrability of NLEEs in 3+1
dimensions is still not a well-solved problem~\cite{a08}) and
admit more abundant soliton structures~\cite{a09}. It is mentioned
that by dimensional splitting the higher-dimensional problem can
be reduced to several lower-dimensional ones which are easier to
treat with the available tools~\cite{a09aa,a09a}. In recent
studies, many (2+1)-dimensional integrable NLEEs have shown to be
relevant with some known (1+1)-dimensional soliton equations by
the nonlinearization of their Lax pairs and adjoint Lax
pairs~\cite{a09a,a10,a11,a12}. Through such a decomposition, a
submanifold of solutions for the given (2+1)-dimensional
integrable NLEE is obtainable by solving the resulting
(1+1)-dimensional integrable systems.

The modified Kadomtsev-Petviashvili (mKP) equation~\cite{a19}, as
one of the most important integrable NLEEs in 2+1 dimensions, has
been derived in many physical applications such as the propagation
of ion-acoustic waves in a plasma with non-isothermal
electrons~\cite{a20} and the electromagnetic wave description in
an isotropic charge-free infinite ferromagnetic thin
film~\cite{a21}. It has been found that the mKP equation is able
to be decomposed into the first two nontrivial nonlinear systems
in the Burgers hierarchy~\cite{a12}, two-coupled Korteweg-de Vries
(KdV) hierarchy~\cite{a15}, two-coupled Chen-Lee-Liu (CLL)
hierarchy~\cite{a12,a21a}, two-coupled KN hierarchy~\cite{a22,a23}
and some other soliton hierarchies~\cite{a13}. But to our
knowledge none has given a systematic way of determining its all
decompositions to two (1+1)-dimensional integrable NLEEs in the
same hierarchy, which means that some new or more generalized
integrable decompositions for the mKP equation have not been
uncovered as yet.

In Ref.~\cite{a12}, the authors have pointed out that the
following mKP equation
\renewcommand{\theequation}{1.\arabic{equation}}
\setcounter{equation}{0}
\begin{equation}
q_{t}=\frac{1 }{4}\left(q_{xxx} - 6\,q^{2}q_{x} -
6\,q_{x}\,\partial_{x}^{-1}q_{y} +3\,\partial_{x}^{-1}q_{yy}
\right),
 \label{eqa01}
\end{equation}
can be  constrained into the two-coupled CLL and high-order CLL
systems by imposing the nonlinearization on  both the associated
Lax pair and another auxillary Lax pair. The present paper is
intended to make a further investigation on Eqn.~(\ref{eqa01}) by
proposing two types of generalized integrable decompositions which
respectively reduce Eqn.~(\ref{eqa01}) to the first two nontrivial
members in the $ 2N $-coupled CLL and KN hierarchies.  On this
basis, our next concern is to derive the Lax representations of
the decomposed (1+1)-dimensional nonlinear systems and construct
their respective Darboux transformations by which some new
solitary-wave solutions are expected to be revealed for
Eqn.~(\ref{eqa01}).

\vspace{7mm}

\noindent {\textbf{2. Proposal of generalized integrable
decompositions}}

As indicated in Refs.~\cite{a12,a19},  Eqn.~(\ref{eqa01}) is
associated with the following linear system
\renewcommand{\theequation}{2.\arabic{equation}}
\setcounter{equation}{0}
\begin{subequations}
\begin{align}
&  u_{y}  = L_{1}\,u, \ \  L_{1} = \partial_{x}^{2} - 2\,q\, \partial_{x}, \label{eqa02a}  \\
&   u_{t} = M_{1}\,u,  \ \ M_{1} =
\partial_{x}^{3}-3\,q\,\partial_{x}^{2} + \frac{3}{2}\left(q^2 -
q_{x}
 - \partial_{x}^{-1}q_{y}\right)\,\partial_{x}, \label{eqa02b}
\end{align}  \label{eqa02}
\end{subequations}
\hspace{-1.5mm}from which we follow the definition of the adjoint
of a differential operator (for a differential operator in the
form $ \Omega=\Sigma a_{k}\,\partial_{x}^{k}$, its adjoint form is
taken as $ \Omega^{*} = \Sigma (-\partial_{x})^{k}\,a_{k} $, where
the asterisk denotes the adjoint operator~\cite{a10}) and obtain
other three linear systems respectively with respect to $ v =
(u_{x})^{*} $, $ m = u^{*} $ and $ p = v^{*} $, as follows:
\begin{subequations}
\begin{align}
& \hspace{-13.5mm}
v_{y} = L_{2}\,v, \ \ \  L_{2} = -\,\partial_{x}^{2} - 2\,q\, \partial_{x}, \label{eqa03a}  \\
&  \hspace{-13.5mm} v_{t} = M_{2}\,v,  \ \ M_{2} =
\partial_{x}^{3} + 3\,q\,\partial_{x}^{2} + \frac{3}{2}\left(q^2 +
q_{x} -
\partial_{x}^{-1}q_{y}\right) \partial_{x},  \label{eqa03b}
\end{align}  \label{eqa03}
\end{subequations}
\vspace{-8mm}
\begin{subequations}
\begin{align}
&
m_{y}  = -L_{1}^{*}\,m, \ \ \ L_{1}^{*}= \partial_{x}^{2} + 2\,q\, \partial_{x}+ 2\,q_{x}, \label{eqa04a}  \\
& \nonumber m_{t} = -M_{1}^{*}\,m,  \ \ M_{1}^{*}=
-\,\partial_{x}^{3} - 3\,q\,\partial_{x}^{2} -
\frac{3}{2}\left(q^2 + 3\,q_{x} -
\partial_{x}^{-1}q_{y}\right)\partial_{x}  \\
&  \hspace{39.5mm} + \frac{3}{2} \left(q_{y}-2\,q\, q_{x} -
q_{xx}\right), \label{eqa04b}
\end{align}  \label{eqa04}
\end{subequations}
\vspace{-8mm}
\begin{subequations}
\begin{align}
& \hspace{-2mm}
p_{y}  = -L_{2}^{*}\,p, \ \ \ L_{2}^{*}= - \,\partial_{x}^{2} + 2\,q\, \partial_{x}+ 2\,q_{x}, \label{eqa05a}  \\
&  \hspace{-2mm} \nonumber p_{t} = -M_{2}^{*}\,p,  \ \ M_{2}^{*}=
-\,\partial_{x}^{3} + 3\,q\,\partial_{x}^{2} -
\frac{3}{2}\left(q^2 - 3\,q_{x} -
\partial_{x}^{-1}q_{y}\right) \partial_{x} \\
& \hspace{34.5mm} + \frac{3}{2} \left(q_{y}-2\,q\, q_{x} +
q_{xx}\right), \label{eqa05b}
\end{align}  \label{eqa05}
\end{subequations}
\hspace{-2mm}where System~(\ref{eqa03}) has also been given in
Ref.~\cite{a12}, while Systems~(\ref{eqa04}) and~(\ref{eqa05}) are
exhibited here for the first time. By direct calculations, we find
that the compatibility conditions $ v_{yt} = v_{ty} $, $ m_{yt} =
m_{ty} $ and $ p_{yt} = p_{ty} $ all give rise to
Eqn.~(\ref{eqa01}), which suggests that
Systems~(\ref{eqa03})--(\ref{eqa05}) are other three different Lax
pairs of Eqn.~(\ref{eqa01}).

Assuming $ u_{j} $ and $ v_{j} $ ($ j=1,2,\ldots ,N$) respectively
satisfy Systems~(\ref{eqa02}) and (\ref{eqa03}),  we introduce the
following potential constraint~\cite{a10}
\begin{equation}
q_{I} = -\frac{1}{2}\,\sum_{j = 1}^{N}u_{j}\,v_{j},
 \label{eqa06}
\end{equation}
into Systems~(\ref{eqa02}) and (\ref{eqa03}), and obtain the
following $ 2N $-coupled CLL system~\cite{a23a},
\begin{equation}
u_{j,y}-u_{j,xx} - \sum_{k = 1}^{N}\,u_{k}\,v_{k}\, u_{j,x} = 0, \
\ \ \ v_{j,y}  + v_{j,xx} - \sum_{k = 1}^{N}\,v_{k}\,u_{k}\,
v_{j,x} = 0, \ \ (j=1,2,\ldots, N),  \label{eqa07}
\end{equation}
and its high-order generalization
\begin{subequations}
\begin{align}
& \nonumber u_{j,t}-u_{j,xxx} - \frac{3}{2}\,\sum_{k =
1}^{N}\,u_{k}\,v_{k}\, u_{j,xx} - \frac{3}{4}\left[\left(\sum_{k =
1}^{N}\,u_{k}\,v_{k} \right)^{2} + 2\,\sum_{k =
1}^{N}\,u_{k,x}\,v_{k}  \right]u_{j,x}= 0,
\\
&  \hspace{10.5cm} (j=1,2,\ldots, N), \\
& \nonumber
 v_{j,t}-v_{j,xxx} + \frac{3}{2}\,\sum_{k =
1}^{N}\,v_{k}\,u_{k}\, v_{j,xx} - \frac{3}{4}\left[\left(\sum_{k =
1}^{N}\,v_{k}\,u_{k} \right)^{2} - 2\,\sum_{k =
1}^{N}\,v_{k,x}\,u_{k}  \right]v_{j,x}= 0, \\
&  \hspace{10.5cm} (j=1,2,\ldots, N).
\end{align}  \label{eqa08}
\end{subequations}
\hspace{-1.5mm}Similarly, if we constrain the potential as
\begin{equation}
q_{II} = -\frac{1}{2}\,\sum_{j = 1}^{N}m_{j}\,p_{j},
 \label{eqa09}
\end{equation}
where $ m_{j} $ and $ p_{j} $ ($ j=1,2,\ldots ,N$) satisfy
Systems~(\ref{eqa04}) and~(\ref{eqa05}),  respectively, then
Eqns.~(\ref{eqa04a}) and~(\ref{eqa05a}) are nonlinearized  into
the $ 2N $-coupled KN system~\cite{a05},
\begin{eqnarray}
& & \nonumber \hspace{-1mm} m_{j,y} + m_{j,xx} - \left(\sum_{k =
1}^{N}\,m_{k}\,p_{k}\,m_{j} \right)_{x} = 0, \ \ \ \ p_{j,y} -
p_{j,xx} - \left(\sum_{k = 1}^{N}\,p_{k}\,m_{k}\,p_{j} \right)_{x}
= 0, \\
& & \hspace{10.2cm} \ \ (j=1,2,\ldots, N), \label{eqa10}
\end{eqnarray}
and Eqns.~(\ref{eqa04b}) and~(\ref{eqa05b}) become
\begin{subequations}
\begin{align}
& \nonumber m_{j,t}-m_{j,xxx}- \frac{3}{2}\,\left[\left(\sum_{k =
1}^{N}\,m_{k}\,p_{k} \right)^{2} m_{j} - \sum_{k =
1}^{N}\,m_{k}\,p_{k}\,m_{j,x}- \sum_{k =
1}^{N}\,m_{k,x}\,p_{k}\,m_{j}  \right]_{x}= 0,
\\
&  \hspace{11.1cm} (j=1,2,\ldots, N), \\
& \nonumber p_{j,t}-p_{j,xxx}- \frac{3}{2}\,\left[\left(\sum_{k =
1}^{N}\,p_{k}\,m_{k} \right)^{2} p_{j} + \sum_{k =
1}^{N}\,p_{k}\,m_{k}\,p_{j,x} + \sum_{k =
1}^{N}\,p_{k,x}\,m_{k}\,p_{j} \right]_{x}= 0, \\
&  \hspace{10.4cm} (j=1,2,\ldots, N),
\end{align}  \label{eqa11}
\end{subequations}
\hspace{-2mm}which is a generalized high-order version of
System~(\ref{eqa10}). Without any difficulty, one can check that
the above two constrained potentials $ q_{I} $ and $ q_{II} $ both
satisfy Eqn.~(\ref{eqa01}) exactly. Thus, we have got two
generalized integrable decompositions for Eqn.~(\ref{eqa01}), as
follows:

\vspace{3mm}

  \textbf{Decomposition I}: If $ (u_{j},v_{j}) $ ($
j=1,2,\ldots ,N$) is a compatible solution of
Systems~(\ref{eqa07}) and~(\ref{eqa08}), then the function  $
q_{I} $ determined by Expression~(\ref{eqa06}) solves the mKP
equation~(\ref{eqa01}).

 \textbf{Decomposition II}: If $ (m_{j},p_{j}) $ ($
j=1,2,\ldots ,N$) is a compatible solution of
Systems~(\ref{eqa10}) and~(\ref{eqa11}), then the function  $
q_{II} $ determined by Expression~(\ref{eqa09}) solves the mKP
equation~(\ref{eqa01}).

Note that the two proposed decompositions in Refs.~\cite{a12,a21a}
and Refs.~\cite{a22,a23}, respectively, correspond to the special
cases of Decompositions I and II when $ N=1 $. Moreover, the
authors in Refs.~\cite{a22,a23} have also not shown the
association of the decomposition there with the Lax pairs for the
mKP equation.

Under the potential constraints~(\ref{eqa06}) and~(\ref{eqa09}), we
can gain much information about the mKP equation~(\ref{eqa01}) from
Systems~(\ref{eqa07})--(\ref{eqa08}) and
(\ref{eqa10})--(\ref{eqa11}) by means of various known effective
approaches. In the following, to explore more unrevealed solutions
(especially the solitary-wave solutions) of Eqn.~(\ref{eqa01}), we
will deal with the decomposed (1+1)-dimensional nonlinear systems by
employing the Darboux transformation method which has been proved to
be an excellent technique for analytically studying integrable NLEEs
and soliton problems~\cite{a24} in that it gives the general
procedure to recursively generate a series of explicit solutions
including the multi-soliton solutions from an initial
solution~\cite{a25}. Once the Darboux transformation for a given
NLEE is constructed, one only needs to solve a linear differential
system (i.e., the Lax pair with an initial potential) and perform
tedious but not complicated algebraic operations~\cite{a26}. Hereby,
an obvious advantage of the Darboux transformation lies in that the
iterative algorithm is purely algebraic and very computerizable by
virtue of symbolic computation~\cite{a26a,a26b,n3}.

\vspace{7mm}

 \noindent {\textbf{3. Lax representations and Darboux transformations of Systems~(\ref{eqa07}) and~(\ref{eqa08})}}

It is possible that an integrable nonlinear system could be
associated with several linear spectral problems, which might lead
to different Darboux transformations. We consider the $
(N\!+\!1)\times(N\!+\!1) $ linear eigenvalue problem and
interestingly find that Systems~(\ref{eqa07}) and~(\ref{eqa08})
admit two different kinds of Lax representations, in which the
first one is of the form
\renewcommand{\theequation}{3.\arabic{equation}}
\setcounter{equation}{0}
\begin{subequations}
\begin{align}
& \mathit{\Psi}_{x}= U^{(1)} \mathit{\Psi} = \big[
\lambda\,U^{(1)}_{0} +
U^{(1)}_{1} \big]\!\mathit{\Psi},  \label{eqa12} \\
& \mathit{\Psi}_{y}= V^{(1)}\mathit{\Psi} =  \big[
\lambda^{2}\,V^{(1)}_{0} +
\lambda\,V^{(1)}_{1}+ V^{(1)}_{2} \big]\!\mathit{\Psi}, \label{eqa13} \\
& \mathit{\Psi}_{t}= W^{(1)}\mathit{\Psi} = \big[
\lambda^{3}\,W^{(1)}_{0} + \lambda^{2}\,W^{(1)}_{1} +
\lambda\,W^{(1)}_{2} + W^{(1)}_{3} \big]\!\mathit{\Psi},
\label{eqa13c}
\end{align} \label{eqa14}
\end{subequations}
\hspace{-1.8mm}where $ \lambda $ is the eigenvalue parameter, $
\mathit{\Psi}\! =\! (\psi_{1}, \psi_{2}, \ldots, \psi_{N+1})^{T} $
(the superscript $ T $ denotes the vector transpose) is the vector
eigenfunction,  the matrices $ U^{(1)}_{i} $, $ V^{(1)}_{k} $ and
$ W^{(1)}_{l} $ ($ i=0,1 $; $ k=0,1,2 $; $ l=0,1,2,3$) are
expressible in the form
\begin{eqnarray}
& & \hspace{-5mm} V^{(1)}_{0} = 2\,U^{(1)}_{0}, \ \ \ \
W^{(1)}_{0} = 4\,U^{(1)}_{0}, \ \ \ \ W^{(1)}_{1} =
 2\,V^{(1)}_{1}, \label{eqa15} \\
& & \hspace{-5mm}
U^{(1)}_{0} = \begin{pmatrix} 1 & O_{1}^{T} \\
V & -I  \end{pmatrix}, \ \ \ \
 U^{(1)}_{1} =
\begin{pmatrix} 0 & -U \\
O_{1} & -\frac{1}{2}\,V U
\end{pmatrix}, \label{eqa16}
\end{eqnarray}
\begin{eqnarray}
& & \hspace{-14mm} V^{(1)}_{1} =
\begin{pmatrix} UV & -2\,U \\
\frac{1}{2}\,V U V-V_{x} & -VU
\end{pmatrix}, \ \ \ \
V^{(1)}_{2} =
\begin{pmatrix} 0 & -\frac{1}{2}\,UVU - U_{x}   \\
O_{1} & -\frac{1}{4}\,V U V U + \frac{1}{2} \left(V_{x}U - VU_{x}
\right)
\end{pmatrix}, \label{eqa17} \\
& & \hspace{-14mm} W^{(1)}_{2} =
\begin{pmatrix} \frac{1}{2}\,A + U_{x}V - UV_{x} & -U V U-2\,U_{x} \\
\frac{1}{4}\,V A + \frac{1}{2}\,V\left(U_{x}V - UV_{x} \right) -
V_{x} U V + V_{xx}  & -\frac{1}{2}\,V U V U + V_{x}U - VU_{x}
\end{pmatrix},  \label{eqa18} \\
 & & \nonumber \hspace{-14mm}
 W^{(1)}_{3} = \frac{1}{2} \times \\
& & \hspace{-14mm}
\begin{pmatrix} 0 &  -\frac{1}{2}\,AU + \left(U V_{x} -
U_{x} V \right)U -2\,UVU_{x} - 2\,U_{xx} \\
O_{1} & -\frac{1}{4}\,V A U + V_{x} U V U - VUVU_{x}  +
\frac{1}{2}\,V \!\left(UV_{x} - U_{x}V\right)U + V_{x}U_{x}
-V_{xx}U - VU_{xx}
\end{pmatrix}, \label{eqa19}
\end{eqnarray}
with  $ I $ as the $ N \times N $ identity matrix, $ A= UVUV $, $
U= \left(u_{1},u_{2},\ldots,u_{N} \right) $, $ V =
\left(v_{1},v_{2},\ldots,v_{N} \right)^{T} $ and $ O_{1}=\left(0,
0, \ldots, 0 \right)^{T} $.  Based on the matrix-form inverse
scattering formulation  in Ref.~\cite{a27}, the second Lax
representation of Systems~(\ref{eqa07}) and~(\ref{eqa08}) is
presented as follows:
\begin{subequations}
\begin{align}
& \hspace{-3mm} \mathit{\Psi}_{x}= U^{(2)}\mathit{\Psi} =
\big[\lambda^{2}\,U^{(2)}_{0} +
\lambda\,U^{(2)} _{1} + U^{(2)}_{2} \big ]\!\mathit{\Psi},  \label{eqa20} \\
& \hspace{-3mm} \mathit{\Psi}_{y}= V^{(2)}\mathit{\Psi} =\big[
\lambda^{4}\,V^{(2)}_{0} + \lambda^{3}\,V^{(2)}_{1} +
\lambda^{2}\,V^{(2)} _{2} +
\lambda\,V^{(2)}_{3}+ V^{(2)}_{4} \big]\!\mathit{\Psi}, \label{eqa21} \\
& \hspace{-3mm} \mathit{\Psi}_{t}= W^{(2)}\mathit{\Psi} =
\big[\lambda^{6}\,W^{(2)}_{0} +\lambda^{5}\,W^{(2)}_{1}
+\lambda^{4}\,W^{(2)}_{2} + \lambda^{3}\,W^{(2)}_{3} +
\lambda^{2}\,W^{(2)}_{4} + \lambda\,W^{(2)}_{5} + W^{(2)}_{6}
\big]\!\mathit{\Psi},  \label{eqa21c}
\end{align} \label{eqa22}
\end{subequations}
\hspace{-2mm}with
\begin{eqnarray}
& & \hspace{-16.5mm} V^{(2)}_{0} = -2\,U^{(2)}_{0}, \ \ \ \
W^{(2)}_{0} = 4\,U^{(2)}_{0}, \ \ \ \ V^{(2)}_{1} =
 -2\,U^{(2)}_{1}, \label{eqa23} \\
& & \hspace{-16.5mm} W^{(2)}_{1} =  4\,U^{(2)}_{1}, \ \ \ \
W^{(2)}_{2} =
 -2\,V^{(2)}_{2}, \ \ \ \ W^{(2)}_{3} =
 -2\,V^{(2)}_{3}, \label{eqa24} \\
 & &
 \hspace{-16.5mm}
U^{(2)}_{0} = \begin{pmatrix} -1 & O_{1}^{T} \\
O_{1} & I  \end{pmatrix}, \ \
 U^{(2)}_{1} =
\begin{pmatrix} 0 & U \\
V & O_{2}
\end{pmatrix}, \ \
U^{(2)}_{2} =
\begin{pmatrix} 0 & O_{1}^{T} \\
O_{1} & -\frac{1}{2}VU
\end{pmatrix}, \label{eqa25} \\
& & \hspace{-16.5mm} V^{(2)}_{2}=
\begin{pmatrix}
-UV & O_{1}^{T} \\
O_{1} & V U
\end{pmatrix}, \ \ \ \  V^{(2)}_{3}=
\begin{pmatrix}
0 & \frac{1}{2}\,U V U + U_{x} \\
\frac{1}{2}\,V U V - V_{x} & O_{2}
\end{pmatrix}, \label{eqa26} \\
& & \hspace{-16.5mm}
 V^{(2)}_{4}=
\begin{pmatrix}
0 & O_{1}^{T} \\
O_{1} & -\frac{1}{4}\,V U V U + \frac{1}{2}
\left(V_{x}U-VU_{x}\right)
\end{pmatrix},  \label{eqa27}  \\
& & \hspace{-16.5mm} W^{(2)}_{4}=
\begin{pmatrix}
-\frac{1}{2}\,A + UV_{x}-U_{x}V  & O_{1}^{T} \\ O_{1} &
\frac{1}{2}\,V U V U + VU_{x}-V_{x}U
\end{pmatrix},  \label{eqa28} \\
& & \hspace{-16.8mm} \nonumber
 W^{(2)}_{5}= \\
 & & \hspace{-17mm}
\begin{pmatrix}
0 & \frac{1}{4}\,A U + \frac{1}{2}\left(U_{x}V- UV_{x} \right)U+ UVU_{x} + U_{xx} \\
 \frac{1}{4}\,V A - \frac{1}{2}\,V\!\left(UV_{x}-U_{x}V \right) - V_{x}UV + V_{xx} & O_{2}
\end{pmatrix}\! ,   \label{eqa29}
\\ & & \hspace{-16.5mm} \nonumber
W^{(2)}_{6} =  \frac{1}{2} \times \\
& & \hspace{-16mm}
\begin{pmatrix} 0 & O_{1}^{T} \\
O_{1} & -\frac{1}{4}\,V A U - VU_{xx}- V_{xx}U  + V_{x}U_{x} -
VUVU_{x} + V_{x}UVU  + \frac{1}{2}\,V\!\left(UV_{x}-U_{x}V
\right)U
\end{pmatrix},  \label{eqa30}
\end{eqnarray}
where $ O_{2} $ is the $ N \times N $ zero matrix, $ A $, $ I $, $
U $, $ V $ and $ O_{1} $ have been defined as above. Here, it is
easy to verify that Systems~(\ref{eqa07}) and~(\ref{eqa08}) can be
respectively derived from  the zero-curvature conditions $
U^{(i)}_{y} - V^{(i)}_{x} + [U^{(i)}, \, V^{(i)}] = 0 $ and $
U^{(i)}_{t} - W^{(i)}_{x} + [U^{(i)}, \, W^{(i)}] = 0 $ ($ i=1,2
$), where the brackets represent the usual matrix commutator.

It is known that the Darboux transformation is actually a gauge
transformation which relates two different solutions of the same
linear system~\cite{a24}. Starting from System~(\ref{eqa14}), we
construct the first Darboux transformation for
Systems~(\ref{eqa07}) and (\ref{eqa08}) in the  form
\begin{equation}
\mathit{\hat{\Psi}} = \big(\lambda \mathit{\Delta}^{(1)}-
\mathit{\Delta}^{(1)} S^{(1)} \big)\mathit{\Psi}, \ \
\mathit{\Delta}^{(1)}=
\begin{pmatrix}
\mathit{\Delta}_{1}^{(1)} & \mathit{\Delta}_{2}^{(1)} \\
\mathit{\Delta}_{3}^{(1)} & \mathit{\Delta}_{4}^{(1)}
\end{pmatrix}, \ \ S^{(1)} = \begin{pmatrix}
S_{1}^{(1)} & S_{2}^{(1)} \\
S_{3}^{(1)} & S_{4}^{(1)}
\end{pmatrix}, \label{eqa31}
\end{equation}
where $ \mathit{\Delta}_{1}^{(1)} = \delta^{(1)}_{11} $, $
\mathit{\Delta}_{2}^{(1)} = \big( \delta^{(1)}_{12},\ldots,
\delta^{(1)}_{1,N+1} \big ) $,  $ \mathit{\Delta}_{3}^{(1)} =
\big( \delta^{(1)}_{21},\ldots, \delta^{(1)}_{N+1,1} \big)^{T} $,
$ \mathit{\Delta}_{4}^{(1)} = \big( \delta^{(1)}_{ik}\big)_{2\leq
i,\, k \leq N+1} $, $ S_{1}^{(1)} = s^{(1)}_{11} $, $ S_{2}^{(1)}
= \big(  s^{(1)}_{12},\ldots, s^{(1)}_{1,N+1} \big) $, $
S_{3}^{(1)} = \big(  s^{(1)}_{21},\ldots, s^{(1)}_{N+1,1}
\big)^{T} $, $ S_{4}^{(1)} = \big( s^{(1)}_{ik}\big)_{2\leq i,\, k
\leq N+1} $, $  \delta^{(1)}_{ik} $ and $  s^{(1)}_{ik} $ ($ 1\leq
i, k \leq N+1 $) are all the functions of $ x $, $ y $ and $ t $
to be determined, $ \mathit{\hat{\Psi}} $ is required to  satisfy
System~(\ref{eqa14}) with $ U^{(1)} $, $ V^{(1)} $ and $ W^{(1)} $
replaced respectively by $ \hat{U}^{(1)} $, $ \hat{V}^{(1)} $ and
$ \hat{W}^{(1)} $ in which the old potentials $ (u_{j}, v_{j}) $
are transformed into new ones $ (\hat{u}_{j}, \hat{v}_{j}) $ ($
j=1,2,\ldots ,N$).

From the knowledge of the Darboux transformation, we know that the
matrices $ \mathit{\Delta}^{(1)} $, $ S^{(1)} $, $
\hat{U}^{(1)}_{i} $, $ \hat{V}^{(1)}_{k} $ and $ \hat{W}^{(1)}_{l}
$ ($ i=0,1 $; $ k=0,1,2 $; $ l=0,1,2,3 $) must satisfy the
following equations:
\begin{subequations}
\begin{align}
&  \mathit{\Delta}^{(1)}\,U^{(1)}_{0}
-\hat{U}^{(1)}_{0}\mathit{\Delta}^{(1)}
=0, \label{eqa32a} \\
&  \mathit{\Delta}^{(1)}_{x} + \mathit{\Delta}^{(1)}\,U^{(1)}_{1}
- \mathit{\Delta}^{(1)} S^{(1)}\,U^{(1)}_{0} + \hat{U}^{(1)}_{0}
\mathit{\Delta}^{(1)}S^{(1)} -
\hat{U}^{(1)}_{1}\mathit{\Delta}^{(1)} =0,
\label{eqa32b} \\
&  \mathit{\Delta}^{(1)}_{x}S^{(1)} +
\mathit{\Delta}^{(1)}S^{(1)}_{x} +
\mathit{\Delta}^{(1)}S^{(1)}\,U^{(1)}_{1} - \hat{U}^{(1)}_{1}
\mathit{\Delta}^{(1)} S^{(1)}=0,
\label{eqa32c}  \\
&  \mathit{\Delta}^{(1)} V^{(1)}_{1} - \mathit{\Delta}^{(1)}
S^{(1)}\,V^{(1)}_{0} + \hat{V}^{(1)}_{0}
\mathit{\Delta}^{(1)}S^{(1)} -
\hat{V}^{(1)}_{1}\mathit{\Delta}^{(1)} =0, \label{eqa32d} \\
&  \mathit{\Delta}^{(1)}_{y} + \mathit{\Delta}^{(1)} V^{(1)}_{2} -
\mathit{\Delta}^{(1)} S^{(1)}\,V^{(1)}_{1} + \hat{V}^{(1)}_{1}
\mathit{\Delta}^{(1)}S^{(1)} -
\hat{V}^{(1)}_{2}\mathit{\Delta}^{(1)} =0,
\label{eqa32e} \\
&  \mathit{\Delta}^{(1)}_{y}S^{(1)} +
\mathit{\Delta}^{(1)}S^{(1)}_{y} +
\mathit{\Delta}^{(1)}S^{(1)}\,V^{(1)}_{2} - \hat{V}^{(1)}_{2}
\mathit{\Delta}^{(1)}
S^{(1)}=0, \label{eqa32f}   \\
&  \mathit{\Delta}^{(1)} W^{(1)}_{2} - \mathit{\Delta}^{(1)}
S^{(1)}\,W^{(1)}_{1} + \hat{W}^{(1)}_{1}
\mathit{\Delta}^{(1)}S^{(1)} -
\hat{ W}^{(1)}_{2}\mathit{\Delta}^{(1)} =0, \label{eqa32g} \\
&  \mathit{\Delta}^{(1)}_{t} + \mathit{\Delta}^{(1)} W^{(1)}_{3} -
\mathit{\Delta}^{(1)} S^{(1)}\, W^{(1)}_{2} + \hat{W}^{(1)}_{2}
\mathit{\Delta}^{(1)}S^{(1)} -
\hat{W}^{(1)}_{3}\mathit{\Delta}^{(1)} =0,
\label{eqa32h} \\
&  \mathit{\Delta}^{(1)}_{t}S^{(1)} +
\mathit{\Delta}^{(1)}S^{(1)}_{t} + \mathit{\Delta}^{(1)}S^{(1)}\,
W^{(1)}_{3} - \hat{W}^{(1)}_{3} \mathit{\Delta}^{(1)} S^{(1)}=0.
\label{eqa32i}
\end{align} \label{eqa32}
\end{subequations}
\hspace{-1.5mm}By Eqns.~(\ref{eqa32a}) and~(\ref{eqa32b}), we can
directly compute out:
\begin{eqnarray}
& & \mathit{\Delta}_{2}^{(1)}=O_{1}^{T}, \ \  {\mathit{\Delta}_{4,x}^{(1)}} = O_{2},  \label{eqa33} \\
& & \mathit{\delta}_{11,x}^{(1)} =
\mathit{\delta}_{11}^{(1)}S_{2}^{(1)}V-\hat{U}\mathit{\Delta}_{3}^{(1)},
\label{eqa34}
\\ & &
\mathit{\Delta}_{3,x}^{(1)} =2\mathit{\Delta}_{4}^{(1)}S_{3}^{(1)}
+ \mathit{\Delta}_{3}^{(1)}S_{2}^{(1)}V +
\mathit{\Delta}_{4}^{(1)}S_{4}^{(1)}V -
\mathit{\Delta}_{4}^{(1)}VS_{1}^{(1)}-\frac{1}{2}\,\hat{V}\hat{U}\mathit{\Delta}_{3}^{(1)},
\label{eqa35}
\end{eqnarray}
with
\begin{equation}
\hspace{-10mm}   \hat{U} = \big(\mathit{\delta}_{11}^{(1)}U -
2\,\mathit{\delta}_{11}^{(1)}S_{2}^{(1)}\big)\big(\mathit{\Delta}_{4}^{(1)}\big)^{-1},
\ \ \ \hat{V} = \big(2\,\mathit{\Delta}_{3}^{(1)} +
\mathit{\Delta}_{4}^{(1)}V \big)/\mathit{\delta}_{11}^{(1)}.
\label{eqa37}
\end{equation}
Then, using the above results, it can be found that
Eqn.~(\ref{eqa32d}) is lead to be satisfied automatically, while
Eqns.~(\ref{eqa32e}), (\ref{eqa32g}) and~(\ref{eqa32h}) yield the
following constraint conditions on $ \mathit{\delta}_{11}^{(1)} $,
$\mathit{\Delta}_{3}^{(1)} $ and $ \mathit{\Delta}_{4}^{(1)} $ as
\begin{eqnarray}
& & \hspace{-12mm}   {\mathit{\Delta}_{4,y}^{(1)}}=O_{2},  \ \ \ \ {\mathit{\Delta}_{4,t}^{(1)}}=O_{2}, \label{eqa38} \\
& & \hspace{-12mm} \nonumber
 \mathit{\delta}_{11,y}^{(1)} =
2\,\mathit{\delta}_{11}^{(1)}S_{2}^{(1)}VS_{1}^{(1)}-\frac{1}{2}\,\hat{U}\hat{V}\hat{U}\mathit{\Delta}_{3}^{(1)}
-\hat{U}_{x}\mathit{\Delta}_{3}^{(1)} +
\frac{1}{2}\,\mathit{\delta}_{11}^{(1)}S_{2}^{(1)}V U V
-\mathit{\delta}_{11}^{(1)}S_{2}^{(1)}V_{x} \\
& & \hspace{-12mm} \ \ \ \ \ \ \ \ \
+\,2\,\mathit{\delta}_{11}^{(1)}US_{3}^{(1)} -
4\,\mathit{\delta}_{11}^{(1)}S_{2}^{(1)}S_{3}^{(1)}, \label{eqa39} \\
& & \hspace{-12mm} \nonumber    \mathit{\delta}_{11,t}^{(1)} =
\frac{1}{2}\,\mathit{\delta}_{11}^{(1)}S_{2}^{(1)}VU_{x}V
-\,\frac{1}{2}\,\mathit{\delta}_{11}^{(1)}S_{2}^{(1)}VUV_{x}
+\frac{1}{4}\,\mathit{\delta}_{11}^{(1)}S_{2}^{(1)} V A + \hat{U}
\hat{V} \hat{U} \mathit{\Delta}_{4}^{(1)}S_{3}^{(1)}
-\hat{U}_{xx}\mathit{\Delta}_{3}^{(1)}  \\
& & \hspace{-12mm} \nonumber \ \ \ \ \ \ \ \ \ +\,
\mathit{\delta}_{11}^{(1)}\hat{U}\hat{V}_{x}S_{1}^{(1)}-\mathit{\delta}_{11}^{(1)}\hat{U}_{x}\hat{V}S_{1}^{(1)}
 -\frac{1}{2}\,\mathit{\delta}_{11}^{(1)}\hat{A}S_{1}^{(1)} +
\hat{U}\hat{V}\hat{U}\mathit{\Delta}_{3}^{(1)}S_{1}^{(1)} +
2\,\hat{U}_{x}\mathit{\Delta}_{3}^{(1)}S_{1}^{(1)}
\\
& & \hspace{-12mm}   \nonumber \ \ \ \ \ \ \ \ \
 +\,
\mathit{\delta}_{11}^{(1)}S_{1}^{(1)}U_{x}V
-\mathit{\delta}_{11}^{(1)}S_{1}^{(1)}UV_{x}+
\frac{1}{2}\,\mathit{\delta}_{11}^{(1)}S_{1}^{(1)}A -
\mathit{\delta}_{11}^{(1)}S_{2}^{(1)}V_{x} U V +
\mathit{\delta}_{11}^{(1)}S_{2}^{(1)}V_{xx}
\\
& & \hspace{-12mm} \ \ \ \ \ \ \ \ \
+\,\frac{1}{2}\,\hat{U}\hat{V}_{x}\hat{U}\mathit{\Delta}_{3}^{(1)}
-\hat{U}\hat{V}\hat{U}_{x}\mathit{\Delta}_{3}^{(1)} -
\frac{1}{2}\,\hat{U}_{x}\hat{V}\hat{U}\mathit{\Delta}_{3}^{(1)}-
\frac{1}{4}\,\hat{A}\hat{U}\mathit{\Delta}_{3}^{(1)}  +
2\,\hat{U}_{x}\mathit{\Delta}_{4}^{(1)}S_{3}^{(1)}, \label{eqa40}
\\ & & \hspace{-12mm}
\nonumber \mathit{\Delta}_{3,y}^{(1)} =
\frac{1}{2}\,\mathit{\Delta}_{3}^{(1)} S_{2}^{(1)}V U V +
\frac{1}{2}\,\mathit{\Delta}_{4}^{(1)} S_{4}^{(1)}V U V -
4\,\mathit{\Delta}_{3}^{(1)}S_{2}^{(1)}S_{3}^{(1)}
-\mathit{\Delta}_{4}^{(1)} S_{4}^{(1)}V_{x}    \\
& & \hspace{-12mm} \nonumber \ \ \ \ \ \ \ \ \
+\,2\,\mathit{\Delta}_{3}^{(1)}US_{3}^{(1)} +
\mathit{\Delta}_{4}^{(1)}VUS_{3}^{(1)} +
4\,\mathit{\Delta}_{4}^{(1)}S_{3}^{(1)} S_{1}^{(1)} +
2\,\mathit{\Delta}_{3}^{(1)}S_{2}^{(1)}VS_{1}^{(1)}  \\
& & \hspace{-12mm} \nonumber \ \ \ \ \ \ \ \ \ +\,
2\,\mathit{\Delta}_{4}^{(1)}S_{4}^{(1)}VS_{1}^{(1)}-
\frac{1}{2}\,\hat{V}\hat{U}_{x}\mathit{\Delta}_{3}^{(1)}
+\frac{1}{2}\,\hat{V}_{x}\hat{U}\mathit{\Delta}_{3}^{(1)}-
\frac{1}{4}\,\hat{V}\hat{U}\hat{V}\hat{U}\mathit{\Delta}_{3}^{(1)}  \\
& & \hspace{-12mm} \nonumber \ \ \ \ \ \ \ \ \
 + \,\mathit{\Delta}_{4}^{(1)}V_{x}S_{1}^{(1)} - \frac{1}{2}\,\mathit{\Delta}_{4}^{(1)}VUVS_{1}^{(1)} -
\mathit{\Delta}_{3}^{(1)}UVS_{1}^{(1)}-2\,\mathit{\Delta}_{4}^{(1)}V\big(S_{1}^{(1)}\big)^{2}
\\
& & \hspace{-12mm} \ \ \ \ \ \ \ \ \ +\,
\mathit{\Delta}_{3}^{(1)}S_{1}^{(1)}UV  -
2\,\mathit{\Delta}_{4}^{(1)}VS_{2}^{(1)}S_{3}^{(1)} +
\mathit{\Delta}_{4}^{(1)}S_{3}^{(1)}UV - \mathit{\Delta}_{3}^{(1)}
S_{2}^{(1)}V_{x},
 \label{eqa41} \\
& & \nonumber  \hspace{-12mm} \mathit{\Delta}_{3,t}^{(1)} =
\mathit{\Delta}_{3}^{(1)} S_{1}^{(1)}U_{x}V
-\mathit{\Delta}_{3}^{(1)} S_{1}^{(1)}UV_{x}
+\frac{1}{2}\,\mathit{\Delta}_{3}^{(1)} S_{1}^{(1)}A -
\mathit{\Delta}_{4}^{(1)}S_{3}^{(1)}UV_{x}
+ \mathit{\Delta}_{4}^{(1)}S_{3}^{(1)}U_{x}V  \\
& & \hspace{-12mm} \nonumber \ \ \ \ \ \ \ \ \ +\,
\frac{1}{2}\,\hat{V}\hat{U}\hat{V}\hat{U}\mathit{\Delta}_{4}^{(1)}S_{3}^{(1)}
- \mathit{\Delta}_{3}^{(1)}S_{2}^{(1)}V_{x}UV-
\frac{1}{2}\,\mathit{\Delta}_{3}^{(1)} S_{2}^{(1)}VUV_{x}
+ \frac{1}{2}\,\mathit{\Delta}_{3}^{(1)} S_{2}^{(1)}VU_{x}V \\
& & \hspace{-12mm} \nonumber \ \ \ \ \ \ \ \ \
+\,\mathit{\Delta}_{3}^{(1)} S_{2}^{(1)}V_{xx} +
\frac{1}{4}\,\mathit{\Delta}_{4}^{(1)}S_{4}^{(1)}V A +
\mathit{\Delta}_{4}^{(1)} S_{4}^{(1)}V_{xx} +
\hat{V}\hat{U}_{x}\mathit{\Delta}_{3}^{(1)} S_{1}^{(1)}
-\hat{V}_{x}\hat{U}\mathit{\Delta}_{3}^{(1)} S_{1}^{(1)}  \\
& & \hspace{-12mm} \nonumber \ \ \ \ \ \ \ \ \
+\,\frac{1}{2}\,\hat{V}\hat{U}\hat{V}\hat{U}\mathit{\Delta}_{3}^{(1)}
S_{1}^{(1)} + \hat{V}_{x}\hat{U}\hat{V}\mathit{\Delta}_{1}^{(1)}
S_{1}^{(1)} +
\frac{1}{2}\,\hat{V}\hat{U}\hat{V}_{x}\mathit{\Delta}_{1}^{(1)}
S_{1}^{(1)} -
\frac{1}{2}\,\hat{V}\hat{U}_{x}\hat{V}\mathit{\Delta}_{1}^{(1)}
S_{1}^{(1)} \\
& & \hspace{-12mm} \nonumber \ \ \ \ \ \ \ \ \ +\,
\frac{1}{2}\,\hat{V}_{x}\hat{U}_{x}\mathit{\Delta}_{3}^{(1)}
-\frac{1}{4}\,\hat{V} \hat{A}\mathit{\Delta}_{1}^{(1)} S_{1}^{(1)}
-\hat{V}_{xx}\mathit{\Delta}_{1}^{(1)}S_{1}^{(1)}
-\frac{1}{2}\,\hat{V}\hat{U}_{xx}\mathit{\Delta}_{3}^{(1)}
-\frac{1}{2}\,\hat{V}_{xx}\hat{U}\mathit{\Delta}_{3}^{(1)} \\
& & \hspace{-12mm}  \nonumber \ \ \ \ \ \ \ \ \ +\,
\frac{1}{4}\,\mathit{\Delta}_{3}^{(1)} S_{2}^{(1)} VA -
\mathit{\Delta}_{4}^{(1)} S_{4}^{(1)}V_{x} U V
-\frac{1}{2}\,\mathit{\Delta}_{4}^{(1)} S_{4}^{(1)}VUV_{x} +
\frac{1}{2}\,\mathit{\Delta}_{4}^{(1)} S_{4}^{(1)}VU_{x}V  \\
& & \hspace{-12mm}  \nonumber \ \ \ \ \ \ \ \ \ + \,
\frac{1}{4}\,\hat{V}\hat{U}\hat{V}_{x}\hat{U}
\mathit{\Delta}_{3}^{(1)}
-\frac{1}{2}\,\hat{V}\hat{U}\hat{V}\hat{U}_{x}\mathit{\Delta}_{3}^{(1)}
+
\frac{1}{2}\,\hat{V}_{x}\hat{U}\hat{V}\hat{U}\mathit{\Delta}_{3}^{(1)}-
\frac{1}{4}\,\hat{V}\hat{U}_{x}\hat{V}\hat{U}\mathit{\Delta}_{3}^{(1)}
\\
& & \hspace{-12mm}  \ \ \ \ \ \ \ \ \
+\,\hat{V}\hat{U}_{x}\mathit{\Delta}_{4}^{(1)}S_{3}^{(1)}
-\frac{1}{8}\,\hat{V}\hat{A}\hat{U}\mathit{\Delta}_{3}^{(1)}-
\hat{V}_{x}\hat{U}\mathit{\Delta}_{4}^{(1)}S_{3}^{(1)}
+\frac{1}{2}\,\mathit{\Delta}_{4}^{(1)} S_{3}^{(1)}A,
\label{eqa42}
\end{eqnarray}
and three redundant equations:
\begin{eqnarray}
& & \hspace{-26mm} \nonumber
\mathit{\delta}_{11}^{(1)}UVS_{2}^{(1)}-\frac{1}{2}\,\mathit{\delta}_{11}^{(1)}U
VU -2\,\mathit{\delta}_{11}^{(1)}S_{2}^{(1)}VS_{2}^{(1)}  +
\mathit{\delta}_{11}^{(1)}S_{2}^{(1)}V U +
4\,\mathit{\delta}_{11}^{(1)}S_{2}^{(1)}S_{4}^{(1)} \\
& &   \hspace{-16mm}  +\,
\frac{1}{2}\,\hat{U}\hat{V}\hat{U}\mathit{\Delta}_{4}^{(1)}-
2\,\mathit{\delta}_{11}^{(1)}US_{4}^{(1)} +
\hat{U}_{x}\mathit{\Delta}_{4}^{(1)} +
2\,\mathit{\delta}_{11}^{(1)}S_{1}^{(1)}U
-\mathit{\delta}_{11}^{(1)}U_{x}   =0, \label{eqa43}
\end{eqnarray}
\begin{eqnarray}
& & \hspace{-12mm} \nonumber  \mathit{\Delta}_{3}^{(1)}U_{x}V -
\mathit{\Delta}_{3}^{(1)}UV_{x} +
\frac{1}{2}\,\mathit{\Delta}_{3}^{(1)}A -
\mathit{\Delta}_{4}^{(1)}V_{x}UV - \frac{1}{2}\,
\mathit{\Delta}_{4}^{(1)}VUV_{x} + \frac{1}{2}\,
\mathit{\Delta}_{4}^{(1)}VU_{x}V \\
& &  \nonumber \hspace{-6mm} + \,\frac{1}{4}\,
\mathit{\Delta}_{4}^{(1)}VA -
2\,\mathit{\Delta}_{3}^{(1)}S_{1}^{(1)}UV -
2\,\mathit{\Delta}_{4}^{(1)}S_{3}^{(1)}UV -
\mathit{\Delta}_{3}^{(1)}S_{2}^{(1)}V U V
-\frac{1}{4}\,\mathit{\delta}_{11}^{(1)}\hat{V}\hat{A} \\
& &  \nonumber \hspace{-6mm}  +\,
2\,\mathit{\Delta}_{4}^{(1)}S_{4}^{(1)}V_{x} + 2\,
\mathit{\Delta}_{3}^{(1)}S_{2}^{(1)}V_{x} -
2\,\hat{V}\hat{U}\mathit{\Delta}_{3}^{(1)}S_{1}^{(1)} +
\mathit{\delta}_{11}^{(1)}\hat{V}\hat{U}\hat{V}S_{1}^{(1)}
-2\,\mathit{\delta}_{11}^{(1)}\hat{V}_{x}S_{1}^{(1)}  \\
& & \nonumber  \hspace{-6mm} +\, \mathit{\Delta}_{4}^{(1)}V_{xx} +
\hat{V}\hat{U}_{x}\mathit{\Delta}_{3}^{(1)} -
\hat{V}_{x}\hat{U}\mathit{\Delta}_{3}^{(1)} +
\frac{1}{2}\,\hat{V}\hat{U}\hat{V}\hat{U}\mathit{\Delta}_{3}^{(1)}
+ \hat{V}_{x}\hat{U}\hat{V}\mathit{\Delta}_{1}^{(1)} -
\mathit{\delta}_{11}^{(1)}\hat{V}_{xx}  \\
& &   \hspace{-6mm} + \,\frac{1}{2}\,\mathit{\delta}_{11}^{(1)}
\hat{V}\hat{U}\hat{V}_{x} -\frac{1}{2}\,\mathit{\delta}_{11}^{(1)}
\hat{V}\hat{U}_{x}\hat{V} -
\mathit{\Delta}_{4}^{(1)}S_{4}^{(1)}VUV -
2\,\hat{V}\hat{U}\mathit{\Delta}_{4}^{(1)}S_{3}^{(1)} =0,
\label{eqa44}  \\
& & \nonumber  \hspace{-12mm}
\frac{1}{2}\,\mathit{\delta}_{11}^{(1)} U V_{x} U -
\mathit{\delta}_{11}^{(1)}U V U_{x} -
\frac{1}{2}\,\mathit{\delta}_{11}^{(1)}U_{x} V U
-\hat{U}\hat{V}\hat{U}\mathit{\Delta}_{4}^{(1)}S_{4}^{(1)} -2\,
\hat{U}_{x}\mathit{\Delta}_{4}^{(1)}S_{4}^{(1)}
-\mathit{\delta}_{11}^{(1)}\hat{U}\hat{V}_{x}S_{2}^{(1)} \\
& &  \nonumber  \hspace{-6mm}
+\,\mathit{\delta}_{11}^{(1)}\hat{U}_{x}\hat{V}S_{2}^{(1)} -
\mathit{\delta}_{11}^{(1)}S_{2}^{(1)}V_{x} U  -
\hat{U}\hat{V}\hat{U}\mathit{\Delta}_{3}^{(1)}S_{2}^{(1)}-2\,
\hat{U}_{x}\mathit{\Delta}_{3}^{(1)}S_{2}^{(1)} -
\frac{1}{2}\,\hat{U}\hat{V}_{x}\hat{U}\mathit{\Delta}_{4}^{(1)} \\
& &   \nonumber  \hspace{-6mm} +\,
\hat{U}\hat{V}\hat{U}_{x}\mathit{\Delta}_{4}^{(1)} +
\frac{1}{2}\,\hat{U}_{x}\hat{V}\hat{U}\mathit{\Delta}_{4}^{(1)}
 +
\mathit{\delta}_{11}^{(1)}S_{1}^{(1)}U V U +
\frac{1}{2}\,\mathit{\delta}_{11}^{(1)}S_{2}^{(1)}V U V U
+ 2\,\mathit{\delta}_{11}^{(1)}S_{1}^{(1)}U_{x} \\
& &   \hspace{-6mm} +\, \mathit{\delta}_{11}^{(1)}S_{2}^{(1)}V
U_{x} + \frac{1}{2}\,\mathit{\delta}_{11}^{(1)}\hat{A}S_{2}^{(1)}
-\mathit{\delta}_{11}^{(1)}U_{xx} +
\hat{U}_{xx}\mathit{\Delta}_{4}^{(1)} -
\frac{1}{4}\,\mathit{\delta}_{11}^{(1)}A U +
\frac{1}{4}\,\hat{A}\hat{U}\mathit{\Delta}_{4}^{(1)}
 = 0, \label{eqa45}
\end{eqnarray}
where $ \hat{A}=\hat{U}\hat{V}\hat{U}\hat{V} $.

The removal of $ \hat{U}^{(1)}_{1} $, $ \hat{V}^{(1)}_{2} $ and $
\hat{W}^{(1)}_{3} $ respectively from Eqns.~(\ref{eqa32c}),
(\ref{eqa32f}) and~(\ref{eqa32i}) by virtue of other six equations
in System~(\ref{eqa32}) gives
\begin{subequations}
\begin{align}
& S_{x}^{(1)} = \big[U_{0}^{(1)}S^{(1)}, S^{(1)} \big] +
\big[U_{1}^{(1)}, S^{(1)} \big], \label{eqa46a} \\
&¡¡S_{y}^{(1)} = \big[V_{0}^{(1)}\big(S^{(1)}\big)^{2},
S^{(1)}\big] + \big[V_{1}^{(1)}S^{(1)}, S^{(1)}\big] +
\big[V_{2}^{(1)}, S^{(1)}\big], \label{eqa46b} \\
& S_{t}^{(1)} = \big[W_{0}^{(1)}\big(S^{(1)}\big)^{3},
S^{(1)}\big] + \big[W_{1}^{(1)}\big(S^{(1)}\big)^{2}, S^{(1)}\big]
+ \big[W_{2}^{(1)}S^{(1)}, S^{(1)}\big] + \big[W_{3}^{(1)},
S^{(1)}\big]. \label{eqa46c}
\end{align} \label{eqa46}
\end{subequations}
\hspace{-1.5mm}Here,  after substitution of
Eqns.~(\ref{eqa34})--(\ref{eqa37}),
Eqns.~(\ref{eqa43})--(\ref{eqa45}) are proved to be comple- tely
covered by Eqn.~(\ref{eqa46a}). In other words,
Eqns.~(\ref{eqa43})--(\ref{eqa45}) are all satisfied identically
with the identity of Eqn.~(\ref{eqa46a}). In a similar way like in
Ref.~\cite{a24}, we  take $ S^{(1)} $ as
\begin{equation}
S^{(1)} = H^{(1)}\Lambda^{(1)}\big(H^{(1)}\big)^{-1},
\label{eqa47}
\end{equation}
with
\begin{equation}
H^{(1)} =\big(h_{1}^{(1)},h_{2}^{(1)},\ldots, h_{N+1}^{(1)}\big),
\ \ \Lambda^{(1)} =
{\rm{diag}}\big(\lambda_{1}^{(1)},\lambda_{2}^{(1)},\ldots,
\lambda_{N+1}^{(1)}\big), \label{eqa48}
\end{equation}
where $ h_{k}^{(1)}=\big(h_{1k}^{(1)}, h_{2k}^{(1)}, \ldots,
h_{N+1,k}^{(1)} \big)^{T} $ corresponds to the column solution of
System~(\ref{eqa14}) with $\lambda = \lambda_{k}^{(1)} $ $\big(
\lambda_{i}^{(1)} \neq \lambda_{k}^{(1)} $ when $ i \neq k $; $ 1
\leq i,k\leq N+1 \big)$, namely,
\begin{subequations}
\begin{align}
& H_{x}^{(1)} = U_{0}^{(1)}H^{(1)}\Lambda^{(1)} + U_{1}^{(1)}H^{(1)}, \label{eqa49a} \\
&¡¡H_{y}^{(1)} = V_{0}^{(1)}H^{(1)}\big(\Lambda^{(1)}\big)^{2} +
V_{1}^{(1)}H^{(1)}\Lambda^{(1)}
+ V_{2}^{(1)}H^{(1)}, \label{eqa49b} \\
& H_{t}^{(1)} = W_{0}^{(1)}H^{(1)}\big(\Lambda^{(1)}\big)^{3} +
W_{1}^{(1)}H^{(1)}\big(\Lambda^{(1)}\big)^{2} +
W_{2}^{(1)}H^{(1)}\Lambda^{(1)} + W_{3}^{(1)}H^{(1)}.
\label{eqa49c}
\end{align} \label{eqa49}
\end{subequations}
\hspace{-1.5mm}By straightforward substitution of
Expression~(\ref{eqa47}) with Eqns.~(\ref{eqa48})
and~(\ref{eqa49}), one can easily verify the identity of
Eqns.~(\ref{eqa32c}), (\ref{eqa32f}) and~(\ref{eqa32i}) (details
ignored). Thus, collecting what have been obtained above, we get
the following:

\textbf{Darboux transformation A}: The matrices  $ \hat{U}^{(1)}
$, $ \hat{V}^{(1)} $ and $ \hat{W}^{(1)} $ have the same forms as
$ U^{(1)} $, $ V^{(1)} $ and $ W^{(1)} $  under the linear
transformation~(\ref{eqa31}) with
Eqns.~(\ref{eqa47})--(\ref{eqa49}), where $
\mathit{\Delta}_{2}^{(1)} = O_{1}^{T} $, $
\mathit{\Delta}_{4}^{(1)} $ is an arbitrary invertible constant
matrix, $ \mathit{\delta}_{11}^{(1)} $ and $
\mathit{\Delta}_{3}^{(1)} $ are determined by Eqns.~(\ref{eqa34}),
(\ref{eqa35}), (\ref{eqa39})--(\ref{eqa42}), and the relationship
between the old and new potentials is constructed by
Transformation~(\ref{eqa37}).

The second Darboux transformation of Systems~(\ref{eqa07}) and
(\ref{eqa08}) based on System~(\ref{eqa22}) is assumed  as the
following form~\cite{a28}
\begin{equation}
 \mathit{\tilde{\Psi}} = \big(\lambda^{2} \mathit{\Delta}^{(2)}-
\lambda \mathit{\Delta}^{(2)} S_{\bot}^{(2)}
-\mathit{\Delta}^{(2)} S_{\top}^{(2)}\big)\mathit{\Psi},
\label{eqa50}
\end{equation}
with
\begin{equation}
\mathit{\Delta}^{(2)}=
\begin{pmatrix}
\mathit{\Delta}_{1}^{(2)} & \mathit{\Delta}_{2}^{(2)} \\
\mathit{\Delta}_{3}^{(2)} & \mathit{\Delta}_{4}^{(2)}
\end{pmatrix}, \ \ S_{\bot}^{(2)} = \begin{pmatrix}
0 & S_{2}^{(2)} \\
S_{3}^{(2)} & O_{2}
\end{pmatrix}, \ \
S_{\top}^{(2)} = \begin{pmatrix}
S_{1}^{(2)} & O_{1}^{T} \\
O_{1} & S_{4}^{(2)}
\end{pmatrix},
\label{eqa51}
\end{equation}
where $ \mathit{\Delta}_{1}^{(2)} = \delta^{(2)}_{11} $, $
\mathit{\Delta}_{2}^{(2)} = \big( \delta^{(2)}_{12},\ldots,
\delta^{(2)}_{1,N+1} \big ) $,  $ \mathit{\Delta}_{3}^{(2)} =
\big( \delta^{(2)}_{21},\ldots, \delta^{(2)}_{N+1,1} \big)^{T} $,
$ \mathit{\Delta}_{4}^{(2)} = \big( \delta^{(2)}_{ik}\big)_{2\leq
i,\, k \leq N+1} $, $ S_{1}^{(2)} = s^{(2)}_{11} $, $ S_{2}^{(2)}
= \big(  s^{(2)}_{12},\ldots, s^{(2)}_{1,N+1} \big) $, $
S_{3}^{(2)} = \big(  s^{(2)}_{21},\ldots, s^{(2)}_{N+1,1}
\big)^{T} $, $ S_{4}^{(2)} = \big( s^{(2)}_{ik}\big)_{2\leq i,\, k
\leq N+1} $, $ \delta^{(2)}_{ik} $ and $  s^{(2)}_{ik} $ ($ 1\leq
i, k \leq N+1 $) are all the functions of $ x $, $ y $ and $ t $
to be determined by a set of equations, as below:
\begin{subequations}
\begin{align}
& \mathit{\Delta}^{(2)}\,U^{(2)}_{0}
-\tilde{U}^{(2)}_{0}\mathit{\Delta}^{(2)} = 0,  \label{eqa52a}  \\
& \mathit{\Delta}^{(2)}\,U^{(2)}_{1} -
\mathit{\Delta}^{(2)}S_{\bot}^{(2)}U^{(2)}_{0} +
\tilde{U}^{(2)}_{0}\mathit{\Delta}^{(2)}S_{\bot}^{(2)} -
\tilde{U}^{(2)}_{1}\mathit{\Delta}^{(2)}
=0,  \label{eqa52b}  \\
&  \nonumber \mathit{\Delta}^{(2)}_{x} +
\mathit{\Delta}^{(2)}\,U^{(2)}_{2} - \mathit{\Delta}^{(2)}
S_{\bot}^{(2)}\,U^{(2)}_{1} - \mathit{\Delta}^{(2)}
S_{\top}^{(2)}\,U^{(2)}_{0} + \tilde{U}^{(2)}_{0}
\mathit{\Delta}^{(2)}S_{\top}^{(2)}   \\
& \hspace{5mm} + \tilde{U}^{(2)}_{1}
\mathit{\Delta}^{(2)}S_{\bot}^{(2)} -
\tilde{U}^{(2)}_{2}\mathit{\Delta}^{(2)} =0,
\label{eqa52c} \\
&  \nonumber  \mathit{\Delta}^{(2)}_{x}S_{\bot}^{(2)} +
\mathit{\Delta}^{(2)}S_{\bot,x}^{(2)} +
\mathit{\Delta}^{(2)}S_{\bot}^{(2)}\,U^{(2)}_{2} +
\mathit{\Delta}^{(2)}S_{\top}^{(2)}\,U^{(2)}_{1} \\
& \hspace{5mm} -
\tilde{U}^{(2)}_{1}\mathit{\Delta}^{(2)}S_{\top}^{(2)} -
\tilde{U}^{(2)}_{2}\mathit{\Delta}^{(2)}S_{\bot}^{(2)}=0,
\label{eqa52d}
\\
&  \mathit{\Delta}^{(2)}_{x}S_{\top}^{(2)} +
\mathit{\Delta}^{(2)}S_{\top,x}^{(2)} +
\mathit{\Delta}^{(2)}S_{\top}^{(2)}\,U^{(2)}_{2} -
\tilde{U}^{(2)}_{2}
\mathit{\Delta}^{(2)} S_{\top}^{(2)}=0,   \label{eqa52e} \\
& \nonumber \mathit{\Delta}^{(2)} V^{(2)}_{2} -
\mathit{\Delta}^{(2)} S_{\bot}^{(2)}\,V^{(2)}_{1} -
\mathit{\Delta}^{(2)} S_{\top}^{(2)}\,V^{(2)}_{0} +
\tilde{V}^{(2)}_{0}
\mathit{\Delta}^{(2)}S_{\top}^{(2)} \\
&  \hspace{5mm} +  \tilde{V}^{(2)}_{1}
\mathit{\Delta}^{(2)}S_{\bot}^{(2)}
- \tilde{V}^{(2)}_{2}\mathit{\Delta}^{(2)} = 0, \label{eqa52f}  \\
& \nonumber \mathit{\Delta}^{(2)} V^{(2)}_{3} -
\mathit{\Delta}^{(2)} S_{\bot}^{(2)}\,V^{(2)}_{2} -
\mathit{\Delta}^{(2)} S_{\top}^{(2)}\,V^{(2)}_{1} +
\tilde{V}^{(2)}_{1}
\mathit{\Delta}^{(2)}S_{\top}^{(2)} \\
&  \hspace{5mm} +  \tilde{V}^{(2)}_{2}
\mathit{\Delta}^{(2)}S_{\bot}^{(2)}
- \tilde{V}^{(2)}_{3}\mathit{\Delta}^{(2)} = 0, \label{eqa52g} \\
& \nonumber \mathit{\Delta}^{(2)}_{y} + \mathit{\Delta}^{(2)}
V^{(2)}_{4} - \mathit{\Delta}^{(2)} S_{\bot}^{(2)}\,V^{(2)}_{3} -
\mathit{\Delta}^{(2)} S_{\top}^{(2)}\,V^{(2)}_{2} +
\tilde{V}^{(2)}_{2}
\mathit{\Delta}^{(2)}S_{\top}^{(2)} \\
&  \hspace{5mm} +  \tilde{V}^{(2)}_{3}
\mathit{\Delta}^{(2)}S_{\bot}^{(2)}
- \tilde{V}^{(2)}_{4}\mathit{\Delta}^{(2)} = 0, \label{eqa52h} \\
&  \nonumber  \mathit{\Delta}^{(2)}_{y}S_{\bot}^{(2)} +
\mathit{\Delta}^{(2)}S_{\bot,y}^{(2)} +
\mathit{\Delta}^{(2)}S_{\bot}^{(2)}\,V^{(2)}_{4} +
\mathit{\Delta}^{(2)}S_{\top}^{(2)}\,V^{(2)}_{3} \\
& \hspace{5mm} -
\tilde{V}^{(2)}_{3}\mathit{\Delta}^{(2)}S_{\top}^{(2)} -
\tilde{V}^{(2)}_{4}\mathit{\Delta}^{(2)}S_{\bot}^{(2)}=0,
\label{eqa52i}  \\
&  \mathit{\Delta}^{(2)}_{y}S_{\top}^{(2)} +
\mathit{\Delta}^{(2)}S^{(2)}_{\top,y} +
\mathit{\Delta}^{(2)}S_{\top}^{(2)}\,V^{(2)}_{4} -
\tilde{V}^{(2)}_{4} \mathit{\Delta}^{(2)}S_{\top}^{(2)}=0,
\label{eqa52j}
\end{align} \label{eqa52}
\end{subequations}
\vspace{-7mm}
$$
 \hspace{-9mm}
\mathit{\Delta}^{(2)} W^{(2)}_{4} - \mathit{\Delta}^{(2)}
S_{\bot}^{(2)}\,W^{(2)}_{3} - \mathit{\Delta}^{(2)}
S_{\top}^{(2)}\,W^{(2)}_{2} + \tilde{W}^{(2)}_{2}
\mathit{\Delta}^{(2)}S_{\top}^{(2)}
$$
$$ \hspace{-42mm} +\,  \tilde{W}^{(2)}_{3} \mathit{\Delta}^{(2)}S_{\bot}^{(2)} -
\tilde{W}^{(2)}_{4}\mathit{\Delta}^{(2)} = 0, \eqno  \rm{(3.36k)}
$$
$$
 \hspace{-9mm}
\mathit{\Delta}^{(2)} W^{(2)}_{5} - \mathit{\Delta}^{(2)}
S_{\bot}^{(2)}\,W^{(2)}_{4} - \mathit{\Delta}^{(2)}
S_{\top}^{(2)}\,W^{(2)}_{3} + \tilde{W}^{(2)}_{3}
\mathit{\Delta}^{(2)}S_{\top}^{(2)}
$$
$$
\hspace{-42mm} + \, \tilde{W}^{(2)}_{4}
\mathit{\Delta}^{(2)}S_{\bot}^{(2)} -
\tilde{W}^{(2)}_{5}\mathit{\Delta}^{(2)} = 0, \eqno \rm{(3.36l)}
$$
$$
\hspace{4mm} \mathit{\Delta}^{(2)}_{t} + \mathit{\Delta}^{(2)}
W^{(2)}_{6} - \mathit{\Delta}^{(2)} S_{\bot}^{(2)}\,W^{(2)}_{5} -
\mathit{\Delta}^{(2)} S_{\top}^{(2)}\,W^{(2)}_{4} +
\tilde{W}^{(2)}_{4} \mathit{\Delta}^{(2)}S_{\top}^{(2)}
$$
$$
\hspace{-42mm} + \, \tilde{W}^{(2)}_{5}
\mathit{\Delta}^{(2)}S_{\bot}^{(2)} -
\tilde{W}^{(2)}_{6}\mathit{\Delta}^{(2)} = 0,  \eqno \rm{(3.36m)}
$$
$$
 \hspace{-19mm}
\mathit{\Delta}^{(2)}_{t}S_{\bot}^{(2)} +
\mathit{\Delta}^{(2)}S_{\bot,t}^{(2)} +
\mathit{\Delta}^{(2)}S_{\bot}^{(2)}\,W^{(2)}_{6} +
\mathit{\Delta}^{(2)}S_{\top}^{(2)}\,W^{(2)}_{5}
$$
$$
\hspace{-35mm} -
\,\tilde{W}^{(2)}_{5}\mathit{\Delta}^{(2)}S_{\top}^{(2)} -
\tilde{W}^{(2)}_{6}\mathit{\Delta}^{(2)}S_{\bot}^{(2)} = 0,
 \eqno \rm{(3.36n)}
$$
$$
 \hspace{-11mm}
 \mathit{\Delta}^{(2)}_{t}S_{\top}^{(2)} +
\mathit{\Delta}^{(2)}S^{(2)}_{\top,t} +
\mathit{\Delta}^{(2)}S_{\top}^{(2)}\, W^{(2)}_{6} -
\tilde{W}^{(2)}_{6} \mathit{\Delta}^{(2)} S_{\top}^{(2)}=0,
 \eqno \rm{(3.36o)}
$$
in which  $ \tilde{U}^{(2)}_{i} $, $ \tilde{V}^{(2)}_{k} $ and $
\tilde{W}^{(2)}_{l} $ are required to have the same forms as $
U^{(2)}_{i} $, $ V^{(2)}_{k} $ and $ W^{(2)}_{l} $ ($ 0 \leq i
\leq 2 $; $ 0 \leq k$ $ \leq 4 $; $ 0 \leq l \leq 6 $) except that
$ (u_{j}, v_{j}) $ are  replaced respectively with $
(\tilde{u}_{j}, \tilde{v}_{j}) $ ($ j=1,2,\ldots ,N$).

From Eqns.~(\ref{eqa52a}) and~(\ref{eqa52b}), we have the
following results:
\begin{eqnarray}
& & \mathit{\Delta}_{2}^{(2)}=O_{1}^{T}, \ \  {\mathit{\Delta}_{3}^{(2)}}=O_{1},  \label{eqa53}  \\
& &  \tilde{U} = \big(\mathit{\delta}_{11}^{(2)}U -
2\,\mathit{\delta}_{11}^{(2)}S_{2}^{(2)}\big)\big(\mathit{\Delta}_{4}^{(2)}\big)^{-1},
\ \ \  \tilde{V} = \big( \mathit{\Delta}_{4}^{(2)}V +
2\,\mathit{\Delta}_{4}^{(2)}S_{3}^{(2)}
\big)/\mathit{\delta}_{11}^{(2)}, \label{eqa55}
\end{eqnarray}
which are further substituted into Eqns.~(\ref{eqa52c}),
(\ref{eqa52f})--(\ref{eqa52h}) and~(3.36k)--(3.36m), determining
that $ \mathit{\delta}_{11}^{(2)} $ and $
\mathit{\Delta}_{4}^{(2)} $ should satisfy the conditions as
\begin{eqnarray}
& & \hspace{-5mm}  \mathit{\Delta}_{4,x}^{(2)}=O_{2}, \ \
\mathit{\Delta}_{4,y}^{(2)}=O_{2}, \ \
\mathit{\Delta}_{4,t}^{(2)}=O_{2}, \label{eqa56}
\\ & &  \hspace{-5mm}
\mathit{\delta}_{11,x}^{(2)} =
\mathit{\delta}_{11}^{(2)}S_{2}^{(2)}V -
\mathit{\delta}_{11}^{(2)}U S_{3}^{(2)} +
2\,\mathit{\delta}_{11}^{(2)}S_{2}^{(2)}S_{3}^{(2)}, \label{eqa57}
\\  & & \hspace{-5mm} \nonumber
\mathit{\delta}_{11,y}^{(2)}=
\tilde{U}\tilde{V}\mathit{\delta}_{11}^{(2)}S_{1}^{(2)} +
\frac{1}{2}\,\mathit{\delta}_{11}^{(2)}S_{2}^{(2)}V U V -
\mathit{\delta}_{11}^{(2)}S_{2}^{(2)}V_{x} -
\mathit{\delta}_{11}^{(2)}S_{1}^{(2)} U V \\
& & \hspace{7.5mm}
-\frac{1}{2}\,\tilde{U}\tilde{V}\tilde{U}\mathit{\Delta}_{4}^{(2)}S_{3}^{(2)}
-\tilde{U}_{x}\mathit{\Delta}_{4}^{(2)}S_{3}^{(2)},  \label{eqa58}
\\ & &  \hspace{-5mm} \nonumber
\mathit{\delta}_{11,t}^{(2)} =
\frac{1}{4}\,\mathit{\delta}_{11}^{(2)}S_{2}^{(2)}V A -
\mathit{\delta}_{11}^{(2)}S_{2}^{(2)}V_{x} U V
-\frac{1}{2}\,\mathit{\delta}_{11}^{(2)}S_{2}^{(2)}V U V_{x} +
\mathit{\delta}_{11}^{(2)}S_{2}^{(2)}V_{xx}   \\
& &  \nonumber \hspace{7.5mm} +\,
\frac{1}{2}\,\mathit{\delta}_{11}^{(2)}S_{2}^{(2)}V U_{x} V -
\frac{1}{2}\, \mathit{\delta}_{11}^{(2)}S_{1}^{(2)} A +
\mathit{\delta}_{11}^{(2)}S_{1}^{(2)}U V_{x} -
\mathit{\delta}_{11}^{(2)}S_{1}^{(2)}U_{x} V \\
& &  \nonumber \hspace{7.5mm} +\,
\frac{1}{2}\,\tilde{U}\tilde{V}_{x}\tilde{U}\mathit{\Delta}_{4}^{(2)}S_{3}^{(2)}
-
\frac{1}{4}\,\tilde{A}\,\tilde{U}\mathit{\Delta}_{4}^{(2)}S_{3}^{(2)}-
\tilde{U}\tilde{V}\tilde{U}_{x}\mathit{\Delta}_{4}^{(2)}S_{3}^{(2)}
 +
\frac{1}{2}\,\mathit{\delta}_{11}^{(2)} \tilde{A} S_{1}^{(2)}
 \\
& &  \hspace{7.5mm} + \, \mathit{\delta}_{11}^{(2)} \tilde{U}_{x}
\tilde{V} S_{1}^{(2)}
-\frac{1}{2}\,\tilde{U}_{x}\tilde{V}\tilde{U}\mathit{\Delta}_{4}^{(2)}S_{3}^{(2)}
 -
\mathit{\delta}_{11}^{(2)}\tilde{U}\tilde{V}_{x} S_{1}^{(2)}  -
\tilde{U}_{xx}\mathit{\Delta}_{4}^{(2)}S_{3}^{(2)}, \label{eqa59}
\end{eqnarray}
with $ \tilde{A}=\tilde{U}\tilde{V}\tilde{U}\tilde{V} $,  and
yielding the following redundant equations:
\begin{eqnarray}
& & \hspace{-8mm} \nonumber
\frac{1}{2}\,\mathit{\delta}_{11}^{(2)}U V U +
\mathit{\delta}_{11}^{(2)}U_{x}
-\frac{1}{2}\,\tilde{U}\tilde{V}\tilde{U}\mathit{\Delta}_{4}^{(2)}
-\tilde{U}_{x}\mathit{\Delta}_{4}^{(2)} -
\tilde{U}\tilde{V}\mathit{\delta}_{11}^{(2)}S_{2}^{(2)} +
2\,\mathit{\delta}_{11}^{(2)}S_{1}^{(2)}U \\
& & \hspace{-2mm} -
2\,\tilde{U}\mathit{\Delta}_{4}^{(2)}S_{4}^{(2)} -
\mathit{\delta}_{11}^{(2)}S_{2}^{(2)}V U = 0,   \label{eqa60} \\
& & \hspace{-8mm}  \nonumber
\frac{1}{2}\,\mathit{\Delta}_{4}^{(2)}V U V -
\mathit{\Delta}_{4}^{(2)}V_{x}
-\frac{1}{2}\,\mathit{\delta}_{11}^{(2)}\tilde{V}
\tilde{U}\tilde{V} + \mathit{\delta}_{11}^{(2)}\tilde{V}_{x} +
\mathit{\Delta}_{4}^{(2)}S_{3}^{(2)}U V +
2\,\mathit{\Delta}_{4}^{(2)}S_{4}^{(2)}V \\
& & \hspace{-2mm} -
2\,\mathit{\delta}_{11}^{(2)}\tilde{V}S_{1}^{(2)} +
\tilde{V}\tilde{U} \mathit{\Delta}_{4}^{(2)}S_{3}^{(2)} = 0,
\label{eqa61}
\end{eqnarray}
\begin{eqnarray}
& & \hspace{-8mm}  \nonumber
\frac{1}{4}\,\mathit{\delta}_{11}^{(2)}A\,U -
\frac{1}{2}\,\mathit{\delta}_{11}^{(2)}U V_{x}U +
\mathit{\delta}_{11}^{(2)}U V U_{x} +
\frac{1}{2}\,\mathit{\delta}_{11}^{(2)}U_{x} V U +
\mathit{\delta}_{11}^{(2)}U_{xx} - \frac{1}{4}\,\tilde{A}
\tilde{U} \mathit{\Delta}_{4}^{(2)}  \\
& & \nonumber \hspace{-2mm} +
\,\frac{1}{2}\,\tilde{U}\tilde{V}_{x}\tilde{U}\mathit{\Delta}_{4}^{(2)}
+ \mathit{\delta}_{11}^{(2)}S_{1}^{(2)}U V U -
\frac{1}{2}\,\tilde{U}_{x}\tilde{V}\tilde{U}\mathit{\Delta}_{4}^{(2)}
-\tilde{U}_{xx}\mathit{\Delta}_{4}^{(2)} -
\mathit{\delta}_{11}^{(2)} S_{2}^{(2)} V U_{x}  \\
& & \nonumber \hspace{-2mm}
+\,\mathit{\delta}_{11}^{(2)}S_{2}^{(2)} V_{x} U -
\tilde{U}\tilde{V}\tilde{U}_{x}\mathit{\Delta}_{4}^{(2)} + 2\,
\mathit{\delta}_{11}^{(2)}S_{1}^{(2)}U_{x} -
\mathit{\delta}_{11}^{(2)}\tilde{U}_{x}\tilde{V}S_{2}^{(2)} -
\frac{1}{2}\,\mathit{\delta}_{11}^{(2)}\tilde{A}S_{2}^{(2)} \\
& &  \nonumber \hspace{-2mm}  +\,
\mathit{\delta}_{11}^{(2)}\tilde{U}\tilde{V}_{x}S_{2}^{(2)}  -
\tilde{U}\tilde{V}\tilde{U}\mathit{\Delta}_{4}^{(2)}S_{4}^{(2)} -
2\,\tilde{U}_{x}\mathit{\Delta}_{4}^{(2)}S_{4}^{(2)}  -
\frac{1}{2}\,\mathit{\delta}_{11}^{(2)}S_{2}^{(2)}V U V U \\
& & \hspace{-2mm}
-\,\tilde{U}\tilde{V}\tilde{U}\mathit{\Delta}_{3}^{(2)} -
2\,\tilde{U}_{x}\mathit{\Delta}_{3}^{(2)} = 0 ,
\label{eqa62}  \\
& &  \hspace{-8mm}  \nonumber
\frac{1}{4}\,\mathit{\Delta}_{4}^{(2)}V A -
\mathit{\Delta}_{4}^{(2)}V_{x} U V -
\frac{1}{2}\,\mathit{\Delta}_{4}^{(2)}V U V_{x}  +
\frac{1}{2}\,\mathit{\Delta}_{4}^{(2)}V U_{x} V +
\mathit{\Delta}_{4}^{(2)}V_{xx}
-\mathit{\delta}_{11}^{(2)}\tilde{V}_{xx}    \\
& & \nonumber \hspace{-2mm}
+\,\mathit{\delta}_{11}^{(2)}\tilde{V}_{x} \tilde{U} \tilde{V} +
\frac{1}{2}\,\mathit{\delta}_{11}^{(2)}\tilde{V}\tilde{U}\tilde{V}_{x}
-
\frac{1}{2}\,\mathit{\delta}_{11}^{(2)}\tilde{V}\tilde{U}_{x}\tilde{V}
+ \frac{1}{2}\,\mathit{\Delta}_{4}^{(2)}S_{3}^{(2)}A  -
\mathit{\Delta}_{4}^{(2)} S_{3}^{(2)}U V_{x}  \\
& & \nonumber \hspace{-2mm}
+\,\mathit{\Delta}_{4}^{(2)}S_{3}^{(2)} U_{x} V -
\frac{1}{4}\,\mathit{\delta}_{11}^{(2)} \tilde{V} \tilde{A}+
\mathit{\Delta}_{4}^{(2)}S_{4}^{(2)}V U V - 2\,
\mathit{\Delta}_{4}^{(2)}S_{4}^{(2)}V_{x} +
2\,\mathit{\delta}_{11}^{(2)}\tilde{V}_{x}S_{1}^{(2)} \\
& & \nonumber  \hspace{-2mm}
+\,\tilde{V}\tilde{U}_{x}\mathit{\Delta}_{3}^{(2)}
-\tilde{V}_{x}\tilde{U}\mathit{\Delta}_{3}^{(2)}+
\frac{1}{2}\,\tilde{V}\tilde{U}\tilde{V}
\tilde{U}\mathit{\Delta}_{3}^{(2)}-
\tilde{V}\tilde{U}\tilde{V}\mathit{\Delta}_{3}^{(2)} +
\frac{1}{2}\,\tilde{V}\tilde{U}\tilde{V}
\tilde{U}\mathit{\Delta}_{4}^{(2)}S_{3}^{(2)}  \\
& &  \hspace{-2mm}  +\, \tilde{V}\tilde{U}_{x}
\mathit{\Delta}_{4}^{(2)}S_{3}^{(2)}-
\tilde{V}_{x}\tilde{U}\mathit{\Delta}_{4}^{(2)}S_{3}^{(2)}
-\mathit{\delta}_{11}^{(2)}\tilde{V}\tilde{U}\tilde{V}S_{1}^{(2)}
+ 2\,\tilde{V}_{x}\mathit{\Delta}_{3}^{(2)} = 0, \label{eqa63}
\end{eqnarray}
where the above four equations,  by substitution of
Eqns.~(\ref{eqa55})--(\ref{eqa58}), can be completely covered by
Eqns.~(\ref{eqa64}) and (\ref{eqa65}) as below.

From Eqns.~(\ref{eqa52d})--(\ref{eqa52e}),
(\ref{eqa52i})--(\ref{eqa52j}) and (3.36n)--(3.36o),  we remove $
\tilde{U}^{(2)}_{1} $, $ \tilde{U}^{(2)}_{2} $, $
\tilde{V}^{(2)}_{3} $, $ \tilde{V}^{(2)}_{4} $, $
\tilde{W}^{(2)}_{5} $ and $ \tilde{W}^{(2)}_{6} $ by use of other
equations in System~(\ref{eqa52}), and obtain
\begin{eqnarray}
& & \hspace{-13mm} \nonumber S_{\bot,x}^{(2)} =
\big[U_{0}^{(2)}\big(S_{\bot}^{(2)}\big)^{2} +
U_{1}^{(2)}S_{\bot}^{(2)} + U_{2}^{(2)}, S_{\bot}^{(2)} \big] +
\big[U_{0}^{(2)}S_{\bot}^{(2)} + U_{1}^{(2)}, S_{\top}^{(2)} \big]
\\
& &  \hspace{0mm} + \big[U_{0}^{(2)}S_{\top}^{(2)}, S_{\bot}^{(2)}
\big] , \label{eqa64}
\\
& & \hspace{-13mm} S_{\top,x}^{(2)} =
\big[U_{0}^{(2)}S_{\bot}^{(2)} + U_{1}^{(2)},
S_{\bot}^{(2)}\big]S_{\top}^{(2)} + \big[U_{0}^{(2)},
S_{\top}^{(2)} \big]S_{\top}^{(2)} + \big[U_{2}^{(2)},
S_{\top}^{(2)} \big], \label{eqa65}
\\ & & \hspace{-13mm} \nonumber
S_{\bot,y}^{(2)} =
\big[V_{0}^{(2)}\big(S_{\top}^{(2)}S_{\bot}^{(2)} +
S_{\bot}^{(2)}S_{\top}^{(2)}\big), S_{\top}^{(2)} \big] +
\big[V_{1}^{(2)}\big(S_{\top}^{(2)}S_{\bot}^{(2)} +
S_{\bot}^{(2)}S_{\top}^{(2)}\big), S_{\bot}^{(2)} \big]
\\
& & \hspace{0mm} \nonumber  +
\big[V_{0}^{(2)}\big(S_{\bot}^{(2)}\big)^{4} + V_{1}^{(2)}\big(
S_{\bot}^{(2)}\big)^{3} + V_{2}^{(2)}\big(S_{\bot}^{(2)}\big)^{2}
+ V_{3}^{(2)}S_{\bot}^{(2)}+ V_{4}^{(2)}, S_{\bot}^{(2)} \big]
\\
& & \hspace{0mm}  \nonumber +
\big[V_{0}^{(2)}\big(S_{\bot}^{(2)}\big)^{2}S_{\top}^{(2)} +
V_{0}^{(2)}S_{\top}^{(2)}\big( S_{\bot}^{(2)}\big)^{2} +
V_{0}^{(2)}S_{\bot}^{(2)}S_{\top}^{(2)}S_{\bot}^{(2)},
S_{\bot}^{(2)} \big]
\\
& & \hspace{0mm} \nonumber + \big[
V_{0}^{(2)}\big(S_{\top}^{(2)}\big)^{2}, S_{\bot}^{(2)} \big] +
\big[ V_{1}^{(2)}S_{\top}^{(2)}, S_{\top}^{(2)}\big] + \big[
V_{2}^{(2)}S_{\top}^{(2)}, S_{\bot}^{(2)}\big]
\\
& & \hspace{0mm} + \big[V_{0}^{(2)}\big(S_{\bot}^{(2)}\big)^{3} +
V_{1}^{(2)}\big( S_{\bot}^{(2)}\big)^{2} +
V_{2}^{(2)}S_{\bot}^{(2)} + V_{3}^{(2)}, S_{\top}^{(2)} \big],
\label{eqa66}
\\ & & \hspace{-13mm} \nonumber  S_{\top,y}^{(2)} =
\big[V_{0}^{(2)}\big(S_{\bot}^{(2)}\big)^{3} + V_{1}^{(2)}\big(
S_{\bot}^{(2)}\big)^{2} + V_{2}^{(2)} S_{\bot}^{(2)} +
V_{3}^{(2)}, S_{\bot}^{(2)} \big]S_{\top}^{(2)} +
\big[V_{4}^{(2)}, S_{\top}^{(2)}\big]
\\
& & \hspace{0mm} \nonumber + \big[V_{0}^{(2)},
S_{\bot}^{(2)}\big]S_{\bot}^{(2)}\big( S_{\top}^{(2)}\big)^{2} +
\big[V_{0}^{(2)}, S_{\bot}^{(2)}\big]S_{\top}^{(2)} S_{\bot}^{(2)}
S_{\top}^{(2)} + \big[V_{2}^{(2)},
S_{\top}^{(2)}\big]S_{\top}^{(2)}
\\
& & \hspace{0mm}  \nonumber  + \big[V_{0}^{(2)},
S_{\top}^{(2)}\big]\big(S_{\bot}^{(2)}\big)^{2}S_{\top}^{(2)} +
\big[V_{0}^{(2)}, S_{\top}^{(2)}\big]\big(S_{\top}^{(2)}\big)^{2}
+ \big[V_{1}^{(2)}, S_{\top}^{(2)}\big]S_{\bot}^{(2)}
S_{\top}^{(2)} \\
& & \hspace{0mm} + \big[V_{1}^{(2)},
S_{\bot}^{(2)}\big]\big(S_{\top}^{(2)}\big)^{2}, \label{eqa67}
\\
& & \hspace{-13mm} \nonumber S_{\bot,t}^{(2)} = \big[W_{6}^{(2)} +
W_{5}^{(2)}S_{\bot}^{(2)}, S_{\bot}^{(2)}\big] + \big[W_{5}^{(2)},
S_{\top}^{(2)}\big]+ \big[W_{4}^{(2)},
S_{\top}^{(2)}\big]S_{\bot}^{(2)} + X_{1}S_{\top}^{(2)}
\\  & &
\hspace{0mm} +\,X_{1}\big(S_{\bot}^{(2)}\big)^{2} +
X_{2}S_{\top}^{(2)}S_{\bot}^{(2)}, \label{eqa68}  \\
& & \hspace{-13mm} S_{\top,t}^{(2)} = \big[W_{6}^{(2)},
S_{\top}^{(2)}\big] + \big[W_{5}^{(2)}, S_{\bot}^{(2)}\big]
S_{\top}^{(2)} + \big[W_{4}^{(2)}, S_{\top}^{(2)}\big]
S_{\top}^{(2)} + X_{1}\big(S_{\bot}^{(2)}\big)^{2}   +\, X_{2}
\big(S_{\top}^{(2)}\big)^{2}, \label{eqa69}
\end{eqnarray}
 where $ X_{1} $ and $ X_{2} $ are defined as
\begin{eqnarray}
& & \hspace{-10mm} \nonumber
 X_{1} =
\big[W_{0}^{(2)}S_{\top}^{(2)}S_{\bot}^{(2)}, S_{\top}^{(2)}\big]
+ \big[W_{0}^{(2)}\big(S_{\bot}^{(2)}\big)^{2}S_{\top}^{(2)},
S_{\bot}^{(2)}\big] +
\big[W_{0}^{(2)}S_{\top}^{(2)}\big(S_{\bot}^{(2)}\big)^{2},
S_{\bot}^{(2)}\big]
\\ & & \hspace{1mm} \nonumber
+ \big[W_{0}^{(2)}S_{\bot}^{(2)}S_{\top}^{(2)},
S_{\top}^{(2)}\big] +
\big[W_{0}^{(2)}\big(S_{\top}^{(2)}\big)^{2}, S_{\bot}^{(2)}\big]
+ \big[W_{0}^{(2)}S_{\bot}^{(2)}S_{\top}^{(2)}S_{\bot}^{(2)},
S_{\bot}^{(2)}\big]
\\
& & \hspace{1mm} \nonumber +
\big[W_{0}^{(2)}\big(S_{\bot}^{(2)}\big)^{3}, S_{\top}^{(2)}\big]
+ \big[W_{0}^{(2)}\big(S_{\bot}^{(2)}\big)^{4},
S_{\bot}^{(2)}\big] +
\big[W_{1}^{(2)}S_{\bot}^{(2)}S_{\top}^{(2)}, S_{\bot}^{(2)}\big]
\\
& & \hspace{1mm} \nonumber +
\big[W_{1}^{(2)}S_{\top}^{(2)}S_{\bot}^{(2)}, S_{\bot}^{(2)}\big]
+ \big[W_{1}^{(2)}\big(S_{\bot}^{(2)}\big)^{3},
S_{\bot}^{(2)}\big] +
\big[W_{1}^{(2)}\big(S_{\bot}^{(2)}\big)^{2}, S_{\top}^{(2)}\big]
\\ & & \hspace{1mm} \nonumber
+ \big[W_{2}^{(2)}S_{\top}^{(2)}, S_{\bot}^{(2)}\big] +
\big[W_{3}^{(2)}, S_{\top}^{(2)}\big] +
\big[W_{3}^{(2)}S_{\bot}^{(2)}, S_{\bot}^{(2)}\big]
+\big[W_{4}^{(2)}, S_{\bot}^{(2)}\big]
\\ & & \hspace{1mm}
 + \big[W_{1}^{(2)}S_{\top}^{(2)},
S_{\top}^{(2)}\big]  +
\big[W_{2}^{(2)}\big(S_{\bot}^{(2)}\big)^{2}, S_{\bot}^{(2)}\big]
+ \big[W_{2}^{(2)}S_{\bot}^{(2)}, S_{\top}^{(2)}\big],
\label{eqa70}
\\
& & \hspace{-10mm} \nonumber
 X_{2} =
\big[W_{0}^{(2)}S_{\top}^{(2)}S_{\bot}^{(2)}, S_{\bot}^{(2)}\big]
+ \big[W_{0}^{(2)}\big(S_{\bot}^{(2)}\big)^{3},
S_{\bot}^{(2)}\big] +
\big[W_{0}^{(2)}S_{\bot}^{(2)}S_{\top}^{(2)}, S_{\bot}^{(2)}\big]
\\ & & \hspace{1mm} \nonumber
+ \big[W_{0}^{(2)}\big(S_{\bot}^{(2)}\big)^{2},
S_{\top}^{(2)}\big] + \big[W_{0}^{(2)}S_{\top}^{(2)},
S_{\top}^{(2)}\big] +
\big[W_{1}^{(2)}\big(S_{\bot}^{(2)}\big)^{2}, S_{\bot}^{(2)}\big]
\\
& & \hspace{1mm} \nonumber +
\big[W_{1}^{(2)}S_{\top}^{(2)},S_{\bot}^{(2)}\big] +
\big[W_{1}^{(2)}S_{\bot}^{(2)}, S_{\top}^{(2)}\big] +
\big[W_{2}^{(2)}S_{\bot}^{(2)}, S_{\bot}^{(2)}\big]
\\ & & \hspace{1mm}
+ \big[W_{2}^{(2)}, S_{\top}^{(2)}\big]
+\big[W_{3}^{(2)},S_{\bot}^{(2)}\big].  \label{eqa71}
\end{eqnarray}
In order to make Eqns.~(\ref{eqa64})--(\ref{eqa69}) satisfied, it
is sufficient to require  the following relation (proof omitted
for brevity):
\begin{equation}
S_{\bot}^{(2)}H^{(2)}\Lambda^{(2)}\big(H^{(2)}\big)^{-1} +
S_{\top}^{(2)} =
H^{(2)}\big(\Lambda^{(2)}\big)^{2}\big(H^{(2)}\big)^{-1},
\label{eqa72}
\end{equation}
with
\begin{equation}
H^{(2)} =\big(h_{1}^{(2)},h_{2}^{(2)},\ldots, h_{N+1}^{(2)}\big),
\ \ \Lambda^{(2)} =
{\rm{diag}}\big(\lambda_{1}^{(2)},\lambda_{2}^{(2)},\ldots,
\lambda_{N+1}^{(2)}\big), \label{eqa73}
\end{equation}
where $ h_{k}^{(2)}=\big(h_{1k}^{(2)}, h_{2k}^{(2)}, \ldots,
h_{N+1,k}^{(2)} \big)^{T} $ is the column solution  of
System~(\ref{eqa22}) with $\lambda = \lambda_{k}^{(2)} $ ($
\lambda_{i}^{(2)} \neq \lambda_{k}^{(2)} $ when $ i \neq k $; $ 1
\leq i,k\leq N+1 $), namely,
\begin{subequations}
\begin{align}
&  \hspace{-5mm} H_{x}^{(2)} =
U_{0}^{(2)}H^{(2)}\big(\Lambda^{(2)}\big)^{2}
+ U_{1}^{(2)}H^{(2)}\Lambda^{(2)}+ U_{2}^{(2)}H^{(2)}, \label{eqa74a} \\
& \hspace{-5mm} \nonumber¡¡H_{y}^{(2)} =
V_{0}^{(2)}H^{(2)}\big(\Lambda^{(2)}\big)^{4} +
V_{1}^{(2)}H^{(2)}\big(\Lambda^{(2)}\big)^{3}  +
V_{2}^{(2)}H^{(2)}\big(\Lambda^{(2)}\big)^{2} \\
& \hspace{7mm} +\,
V_{3}^{(2)}H^{(2)}\Lambda^{(2)}  + V_{4}^{(2)}H^{(2)}, \label{eqa74b} \\
& \hspace{-5mm} \nonumber  H_{t}^{(2)} =
W_{0}^{(2)}H^{(2)}\big(\Lambda^{(2)}\big)^{6} +
W_{1}^{(2)}H^{(2)}\big(\Lambda^{(2)}\big)^{5}  +
W_{2}^{(2)}H^{(2)}\big(\Lambda^{(2)}\big)^{4}   +
W_{3}^{(2)}H^{(2)}\big(\Lambda^{(2)}\big)^{3} \\
&  \hspace{7mm}   + W_{4}^{(2)}H^{(2)}\big(\Lambda^{(2)}\big)^{2}
 + W_{5}^{(2)}H^{(2)}\Lambda^{(2)}
+ W_{6}^{(2)}H^{(2)}. \label{eqa74c}
\end{align} \label{eqa74}
\end{subequations}
\hspace{3mm} \textbf{Darboux transformation B}: The matrices  $
\tilde{U}^{(2)}_{i} $, $ \tilde{V}^{(2)}_{k} $ and $
\tilde{W}^{(2)}_{l} $ have the same forms as $ U^{(2)}_{i} $, $
V^{(2)}_{k} $ and $ W^{(2)}_{l} $ ($ 0 \leq i \leq 2 $; $ 0 \leq
k$ $ \leq 4 $; $ 0 \leq l \leq 6 $) under the linear
transformation~(\ref{eqa50}) with
Eqns.~(\ref{eqa72})--(\ref{eqa74}), where $
\mathit{\Delta}_{2}^{(2)}= O_{1}^{T} $, $
\mathit{\Delta}_{3}^{(2)} = O_{1} $, $ \mathit{\Delta}_{4}^{(2)} $
is an arbitrary invertible constant matrix, $
\mathit{\delta}_{11}^{(2)} $ is determined by
Eqns.~(\ref{eqa57})--(\ref{eqa59}), and the relationship between
the old and new potentials is constructed by
Transformation~(\ref{eqa55}).

\vspace{7mm}

\noindent {\textbf{4. Lax representations and Darboux
transformations of Systems~(\ref{eqa10}) and~(\ref{eqa11})}}

Likewise, we find that Systems~(\ref{eqa10}) and~(\ref{eqa11}) are
also  related to two different types of Lax representations, which
are respectively written as
\renewcommand{\theequation}{4.\arabic{equation}}
\setcounter{equation}{0}
 \vspace{-4mm}
\begin{subequations}
\begin{align}
& \mathit{\Phi}_{x}= U^{(3)} \mathit{\Phi} = \big[
\lambda\,U^{(3)}_{0} +
U^{(3)}_{1}  \big]\!\mathit{\Phi},  \label{eqa75} \\
& \mathit{\Phi}_{y}= V^{(3)}\mathit{\Phi} =  \big[
\lambda^{2}\,V^{(3)}_{0} +
\lambda\,V^{(3)}_{1}+ V^{(3)}_{2}  \big]\!\mathit{\Phi}, \label{eqa76} \\
& \mathit{\Phi}_{t}= W^{(3)}\mathit{\Phi} =  \big[
\lambda^{3}\,W^{(3)}_{0} + \lambda^{2}\,W^{(3)}_{1} +
\lambda\,W^{(3)}_{2} + W^{(3)}_{3}  \big]\!\mathit{\Phi},
\label{eqa76c}
\end{align} \label{eqa77}
\end{subequations}
\hspace{-2mm}with
\begin{eqnarray}
& & \hspace{-12mm} V^{(3)}_{0} = - 2\,U^{(3)}_{0}, \ \ \ \
W^{(3)}_{0} = 4\,U^{(3)}_{0}, \ \ \ \ W^{(3)}_{1} =-
 2\,V^{(3)}_{1}, \label{eqa78} \\
& & \hspace{-12mm}
U^{(3)}_{0} = \begin{pmatrix} 1 & O_{1}^{T} \\
P & -I  \end{pmatrix}, \ \ \ \
 U^{(3)}_{1} =
\begin{pmatrix} 0 & M \\
O_{1} &  O_{2}
\end{pmatrix}, \ \ \ \ V^{(3)}_{1} =
\begin{pmatrix} M P & -2\,M \\
P M P + P_{x} & -P M
\end{pmatrix},   \label{eqa79} \\
& & \hspace{-12mm} V^{(3)}_{2} =
\begin{pmatrix} 0 & M P M - M_{x}   \\
O_{1} & O_{2}
\end{pmatrix},  \  W^{(3)}_{3} =
\begin{pmatrix}0 & \frac{3}{2}\,B M - \frac{3}{2}\,M_{x} P M  - \frac{3}{2}\,M P M_{x}
+ M_{xx} \\
O_{1} & O_{2}
\end{pmatrix},  \label{eqa80}  \\
& & \hspace{-12mm} W^{(3)}_{2} =
\begin{pmatrix} \frac{3}{2}\,B + MP_{x} - M_{x}P & -2MPM + 2 M_{x} \\
\frac{3}{2}\,P B + \frac{3}{2}\,P_{x} M P + \frac{3}{2}\,P M P_{x}
+ P_{xx} & -\frac{3}{2}\,P M P M + P M_{x} - P_{x} M
\end{pmatrix},  \label{eqa81}
\end{eqnarray}
and
\begin{subequations}
\begin{align}
& \hspace{-10mm} \mathit{\Phi}_{x}= U^{(4)}\mathit{\Phi} = \
\big[\lambda^{2}\,U^{(4)}_{0} +
\lambda\,U^{(4)}_{1}  \big]\!\mathit{\Phi},  \label{eqa82} \\
& \hspace{-10mm} \mathit{\Phi}_{y} = V^{(4)}\mathit{\Phi} =  \big[
\lambda^{4}\,V^{(4)}_{0} + \lambda^{3}\,V^{(4)}_{1} +
\lambda^{2}\,V^{(4)}_{2} + \lambda\,V^{(4)}_{3}  \big]\!\mathit{\Phi}, \label{eqa83} \\
& \hspace{-10mm} \mathit{\Phi}_{t}= W^{(4)}\mathit{\Phi} =
\big[\lambda^{6}\,W^{(4)}_{0} +\lambda^{5}\,W^{(4)}_{1}
+\lambda^{4}\,W^{(4)}_{2} + \lambda^{3}\,W^{(4)}_{3} +
\lambda^{2}\,W^{(4)}_{4} + \lambda\,W^{(4)}_{5}
\big]\!\mathit{\Phi}, \label{eqa83c}
\end{align} \label{eqa84}
\end{subequations}
\hspace{-2mm}with
\begin{eqnarray}
& & \hspace{-15mm} V^{(4)}_{0} = 2\,U^{(4)}_{0}, \ \ \ \
W^{(4)}_{0} = 4\,U^{(4)}_{0}, \ \ \ \ V^{(4)}_{1} =
2\,U^{(4)}_{1}, \label{eqa85} \\
& & \hspace{-15mm} W^{(4)}_{1} =  4\,U^{(4)}_{1}, \ \ \ \
W^{(4)}_{2} =
 2\,V^{(4)}_{2}, \ \ \ \ W^{(4)}_{3} =
 2\,V^{(4)}_{3}, \label{eqa86} \\
 & &
 \hspace{-15mm}
U^{(4)}_{0} = \begin{pmatrix} -1 & O_{1}^{T} \\
O_{1} & I  \end{pmatrix}, \ \
 U^{(4)}_{1} =
\begin{pmatrix} 0 & M \\
-P & O_{2}
\end{pmatrix},   \ \ V^{(4)}_{2}=
\begin{pmatrix}
-M P & O_{1}^{T} \\
O_{1} & P M
\end{pmatrix}, \label{eqa87} \\
& & \hspace{-15mm} V^{(4)}_{3}=
\begin{pmatrix}
0 & M P M - M_{x} \\
-P M P - P_{x} & O_{2}
\end{pmatrix}, \label{eqa88}  \\
& & \hspace{-15mm} W^{(4)}_{4}=
\begin{pmatrix}
-\frac{3}{2}\,B + M_{x} P - M P_{x}  & O_{1}^{T} \\ O_{1} &
\frac{3}{2}\,P M P M + P_{x}M - P M_{x}
\end{pmatrix},  \label{eqa89} \\
& & \nonumber \hspace{-15mm}
 W^{(4)}_{5}= \\
& & \hspace{-14mm}
\begin{pmatrix}
0 & \frac{3}{2}\,B M - \frac{3}{2}\,M P M_{x} - \frac{3}{2}\,M_{x} P M + M_{xx} \\
 -\frac{3}{2}\,P B - \frac{3}{2}\,P_{x}M P  - \frac{3}{2}\,P M P_{x} - P_{xx} & O_{2}
\end{pmatrix},   \label{eqa90}
\end{eqnarray}
where  $ \mathit{\Phi}\! =\! (\phi_{1}, \phi_{2}, \ldots,
\phi_{N+1})^{T} $, $ M= \left(m_{1},m_{2},\ldots, m_{N} \right) $,
$ P = \left(p_{1}, p_{2},\ldots, p_{N} \right)^{T} $, $ B= MPMP $,
$ O_{1} $, $ O_{2} $ and $ I $ have the same definitions as in
Section 3. The zero-curvature conditions $ U^{(i)}_{y} -
V^{(i)}_{x} + [U^{(i)}, \, V^{(i)}] = 0 $ and $ U^{(i)}_{t} -
W^{(i)}_{x} + [U^{(i)}, \, W^{(i)}] = 0 $ ($ i=3,4 $) give rise to
Systems~(\ref{eqa10}) and~(\ref{eqa11}), respectively.

Following the procedure in Section 3, we can also arrive at two
Darboux transformations for Systems~(\ref{eqa10})
and~(\ref{eqa11}), as follows:

\textbf{Darboux transformation C}: The linear system~(\ref{eqa77})
is kept invariant by the gauge transformation
\begin{equation}
\mathit{\hat{\Phi}} = \big(\lambda \mathit{\Delta}^{(3)}-
\mathit{\Delta}^{(3)} S^{(3)}\big)\mathit{\Phi}, \ \
\mathit{\Delta}^{(3)}=
\begin{pmatrix}
\mathit{\Delta}_{1}^{(3)} & \mathit{\Delta}_{2}^{(3)} \\
\mathit{\Delta}_{3}^{(3)} & \mathit{\Delta}_{4}^{(3)}
\end{pmatrix}, \ \ S^{(3)} = \begin{pmatrix}
S_{1}^{(3)} & S_{2}^{(3)} \\
S_{3}^{(3)} & S_{4}^{(3)}
\end{pmatrix}, \label{eqa91}
\end{equation}
where $ S^{(3)} = H^{(3)}\Lambda^{(3)}\big(H^{(3)}\big)^{-1} $, $
\Lambda^{(3)}={\rm{diag}}\big(\lambda_{1}^{(3)},\lambda_{2}^{(3)},\ldots,
\lambda_{N+1}^{(3)}\big) $, $
H^{(3)}=\big(h_{1}^{(3)},h_{2}^{(3)},\ldots, h_{N+1}^{(3)}\big) $
with $ h_{k}^{(3)} $  ($ 1 \leq k\leq N+1 $) as the column
solutions of System~(\ref{eqa77}) for  different eigenvalues $
\lambda = \lambda_{k}^{(3)} $, $ \mathit{\Delta}_{2}^{(3)} =
O_{1}^{T} $, $ \mathit{\Delta}_{1}^{(3)} = \delta^{(3)}_{11} $, $
\mathit{\Delta}_{3}^{(3)} = \big( \delta^{(3)}_{21},\ldots,
\delta^{(3)}_{N+1,1} \big)^{T} $, $ \mathit{\Delta}_{4}^{(3)} =
\big( \delta^{(3)}_{ik}\big)_{2\leq i,\, k \leq N+1} $ obey the
following conditions:
\begin{eqnarray}
& &  \hspace{-10mm} \mathit{\delta}_{11,x}^{(3)} =
\hat{M}\mathit{\Delta}_{3}^{(3)} +
\mathit{\delta}_{11}^{(3)}S_{2}^{(3)}P, \label{eqa92}  \\
& &  \nonumber  \hspace{-10mm} \mathit{\delta}_{11,y}^{(3)} =
2\,\hat{M}\mathit{\Delta}_{3}^{(3)}S_{1}^{(3)} -
\mathit{\delta}_{11}^{(3)}\hat{M}\hat{P}S_{1}^{(3)} +
\hat{M}\hat{P}\hat{M}\mathit{\Delta}_{3}^{(3)} -
\hat{M}_{x}\mathit{\Delta}_{3}^{(3)} +
\mathit{\delta}_{11}^{(3)}S_{2}^{(3)}P M P  \\
& &  \hspace{8mm} +\, \mathit{\delta}_{11}^{(3)}S_{2}^{(3)} P_{x}
+ 2\, \hat{M}\mathit{\Delta}_{4}^{(3)}S_{3}^{(3)} +
\mathit{\delta}_{11}^{(3)}S_{1}^{(3)} M P,  \label{eqa93} \\
& &  \nonumber  \hspace{-10mm} \mathit{\delta}_{11,t}^{(3)} =
\mathit{\delta}_{11}^{(3)}\hat{M}_{x}\hat{P}S_{1}^{(3)} -
\mathit{\delta}_{11}^{(3)}\hat{M}\hat{P}_{x}S_{1}^{(3)} -
\frac{3}{2}\,\mathit{\delta}_{11}^{(3)}\hat{B}S_{1}^{(3)} +
2\,\hat{M}\hat{P}\hat{M}\mathit{\Delta}_{3}^{(3)}S_{1}^{(3)}+
\frac{3}{2}\,\hat{B}\hat{M}\mathit{\Delta}_{3}^{(3)}  \\
& &  \nonumber  \hspace{3mm} +\,
2\,\hat{M}\hat{P}\hat{M}\mathit{\Delta}_{4}^{(3)}S_{3}^{(3)}
-2\,\hat{M}_{x}\mathit{\Delta}_{3}^{(3)}S_{1}^{(3)}
-\frac{3}{2}\,\hat{M}\hat{P}\hat{M}_{x}\mathit{\Delta}_{3}^{(3)}
-\frac{3}{2}\,\hat{M}_{x}\hat{P}\hat{M}\mathit{\Delta}_{3}^{(3)}
 \\
& &  \nonumber  \hspace{3mm}
+\,\hat{M}_{xx}\mathit{\Delta}_{3}^{(3)} +
\frac{3}{2}\,\mathit{\delta}_{11}^{(3)}S_{1}^{(3)}B  +
\mathit{\delta}_{11}^{(3)}S_{1}^{(3)}MP_{x} -
\mathit{\delta}_{11}^{(3)}S_{1}^{(3)}M_{x}P -
2\,\hat{M}_{x}\mathit{\Delta}_{4}^{(3)}S_{3}^{(3)}\\
& &  \hspace{3mm} +\,
\frac{3}{2}\,\mathit{\delta}_{11}^{(3)}S_{2}^{(3)}PMP_{x} +
\frac{3}{2}\,\mathit{\delta}_{11}^{(3)}S_{2}^{(3)}P_{x}MP +
\mathit{\delta}_{11}^{(3)}S_{2}^{(3)}P_{xx}  +
\frac{3}{2}\,\mathit{\delta}_{11}^{(3)}S_{2}^{(3)}P B,
\label{eqa94}
\end{eqnarray}
\vspace{-10mm}
\begin{eqnarray}
& & \hspace{-12mm}  \mathit{\Delta}_{3,x}^{(3)}=
2\,\mathit{\Delta}_{4}^{(3)}S_{3}^{(3)} -
\mathit{\Delta}_{4}^{(3)}P S_{1}^{(3)} + \mathit{\Delta}_{3}^{(3)}
S_{2}^{(3)} P + \mathit{\Delta}_{4}^{(3)} S_{4}^{(3)} P,  \label{eqa95} \\
& &  \nonumber \hspace{-12mm} \mathit{\Delta}_{3,y}^{(3)} =
\mathit{\Delta}_{3}^{(3)} S_{2}^{(3)}P M P +
\mathit{\Delta}_{3}^{(3)} S_{2}^{(3)}P_{x} +
\mathit{\Delta}_{4}^{(3)} S_{4}^{(3)}P M P  +
\mathit{\Delta}_{4}^{(3)} S_{4}^{(3)}P_{x} + \hat{P} \hat{M}
\mathit{\Delta}_{3}^{(3)} S_{1}^{(3)} \\
& &  \hspace{1mm} +\,\mathit{\Delta}_{3}^{(3)} S_{1}^{(3)}M P -
\mathit{\delta}_{11}^{(3)}\hat{P} \hat{M} \hat{P}S_{1}^{(3)} +
\mathit{\Delta}_{4}^{(3)} S_{3}^{(3)} M P + \hat{P}
\hat{M}\mathit{\Delta}_{4}^{(3)} S_{3}^{(3)} -
\mathit{\delta}_{11}^{(3)} \hat{P}_{x}S_{1}^{(3)}, \label{eqa96} \\
& &  \nonumber \hspace{-12mm} \mathit{\Delta}_{3,t}^{(3)} =
\frac{3}{2}\,\mathit{\Delta}_{3}^{(3)}S_{1}^{(3)}B +
\frac{3}{2}\,\mathit{\Delta}_{4}^{(3)}S_{3}^{(3)}B +
\mathit{\Delta}_{3}^{(3)}S_{1}^{(3)}M P_{x} -
\mathit{\Delta}_{3}^{(3)}S_{1}^{(3)}M_{x} P  - \mathit{\delta}_{11}^{(3)}\hat{P}_{xx}S_{1}^{(3)}\\
& &  \nonumber \hspace{1mm}  +\,
\mathit{\Delta}_{4}^{(3)}S_{3}^{(3)}M P_{x} +
\frac{3}{2}\,\mathit{\Delta}_{3}^{(3)}S_{2}^{(3)} P B +
\frac{3}{2}\,\mathit{\Delta}_{3}^{(3)}S_{2}^{(3)}P M P_{x} +
\frac{3}{2}\,\mathit{\Delta}_{3}^{(3)}S_{2}^{(3)}P_{x} M P \\
& &  \nonumber \hspace{1mm}  +\,
\mathit{\Delta}_{3}^{(3)}S_{2}^{(3)}P_{xx} +
\frac{3}{2}\,\mathit{\Delta}_{4}^{(3)}S_{4}^{(3)}P B  +
\frac{3}{2}\,\mathit{\Delta}_{4}^{(3)}S_{4}^{(3)}P M P_{x} +
\frac{3}{2}\,\mathit{\Delta}_{4}^{(3)}S_{4}^{(3)}P_{x} M P \\
& &  \nonumber \hspace{1mm}   + \,
\hat{P}_{x}\hat{M}\mathit{\Delta}_{4}^{(3)} S_{3}^{(3)} -
\hat{P}\hat{M}_{x}\mathit{\Delta}_{3}^{(3)}S_{1}^{(3)} +
\hat{P}_{x}\hat{M}\mathit{\Delta}_{3}^{(3)}S_{1}^{(3)} +
\frac{3}{2}\,\hat{P}\hat{M}\hat{P}\hat{M}\mathit{\Delta}_{3}^{(3)}S_{1}^{(3)}
\\
& &  \nonumber \hspace{1mm}  +\,
\frac{3}{2}\hat{P}\hat{M}\hat{P}\hat{M}\mathit{\Delta}_{4}^{(3)}
S_{3}^{(3)}  -
\frac{3}{2}\,\mathit{\delta}_{11}^{(3)}\hat{P}_{x}\hat{M}\hat{P}S_{1}^{(3)}
 -
\mathit{\Delta}_{4}^{(3)}S_{3}^{(3)}M_{x} P  -
\hat{P}\hat{M}_{x}\mathit{\Delta}_{4}^{(3)} S_{3}^{(3)}  \\
& &   \hspace{1mm} +\, \mathit{\Delta}_{4}^{(3)}S_{4}^{(3)}P_{xx}
-
\frac{3}{2}\,\mathit{\delta}_{11}^{(3)}\hat{P}\hat{M}\hat{P}_{x}S_{1}^{(3)}
 -
\frac{3}{2}\,\mathit{\delta}_{11}^{(3)}\hat{P}\hat{B}S_{1}^{(3)},
\label{eqa97}
\end{eqnarray}
\vspace{-10mm}
\begin{eqnarray}
 & &  \hspace{-14mm}  \mathit{\Delta}_{4,x}^{(3)}=
-2\,\mathit{\Delta}_{3}^{(3)}S_{2}^{(3)} -
\mathit{\Delta}_{4}^{(3)}P S_{2}^{(3)}  - \mathit{\Delta}_{3}^{(3)} M, \label{eqa98}  \\
 & &  \nonumber \hspace{-14mm}
 \mathit{\Delta}_{4,y}^{(3)}=
-\mathit{\Delta}_{3}^{(3)}M P M -
2\,\mathit{\Delta}_{3}^{(3)}S_{1}^{(3)} M -
2\,\mathit{\Delta}_{4}^{(3)}S_{3}^{(3)} M -
\mathit{\Delta}_{3}^{(3)}S_{2}^{(3)}P M  -
\mathit{\delta}_{11}^{(3)}\hat{P}_{x}S_{2}^{(3)}  \\
& &  \hspace{-1mm} + \,\hat{P}\hat{M}
\mathit{\Delta}_{3}^{(3)}S_{2}^{(3)}
-\mathit{\Delta}_{4}^{(3)}S_{4}^{(3)}P M  -
\mathit{\delta}_{11}^{(3)}\hat{P} \hat{M} \hat{P}S_{2}^{(3)}  +
\mathit{\Delta}_{3}^{(3)}M_{x}+ \hat{P}\hat{M}
\mathit{\Delta}_{4}^{(3)}S_{4}^{(3)}, \label{eqa99}
\end{eqnarray}
\begin{eqnarray}
& &  \hspace{-14mm} \nonumber \mathit{\Delta}_{4,t}^{(3)}=
\frac{3}{2}\,\mathit{\Delta}_{3}^{(3)}M P M_{x} -
\frac{3}{2}\,\mathit{\Delta}_{3}^{(3)}B M +
\frac{3}{2}\,\mathit{\Delta}_{3}^{(3)}M_{x}PM -
\mathit{\Delta}_{3}^{(3)}M_{xx} +
2\,\mathit{\Delta}_{3}^{(3)}S_{1}^{(3)}M_{x}  \\
& &  \nonumber \hspace{-1mm} +\,
2\,\mathit{\Delta}_{4}^{(3)}S_{3}^{(3)}M_{x} -
2\,\mathit{\Delta}_{3}^{(3)}S_{1}^{(3)}M P M -
2\,\mathit{\Delta}_{4}^{(3)}S_{3}^{(3)}M P M +
\mathit{\Delta}_{3}^{(3)}S_{2}^{(3)}P M_{x}  \\
& &  \nonumber \hspace{-1mm} +
\mathit{\Delta}_{4}^{(3)}S_{4}^{(3)}P M_{x}
-\mathit{\Delta}_{3}^{(3)}S_{2}^{(3)}P_{x} M -
\frac{3}{2}\,\mathit{\Delta}_{3}^{(3)}S_{2}^{(3)}P M P M  -
\mathit{\Delta}_{4}^{(3)}S_{4}^{(3)}P_{x} M  \\
& &  \nonumber \hspace{-1mm} +\, \hat{P}_{x}  \hat{M}
\mathit{\Delta}_{3}^{(3)}S_{2}^{(3)}-
\frac{3}{2}\,\mathit{\Delta}_{4}^{(3)}S_{4}^{(3)}P M P M - \hat{P}
\hat{M}_{x} \mathit{\Delta}_{3}^{(3)}S_{2}^{(3)} -
\frac{3}{2}\,\mathit{\delta}_{11}^{(3)}\hat{P}\hat{B} S_{2}^{(3)}
\\
& &  \nonumber \hspace{-1mm} + \, \frac{3}{2}\,\hat{P} \hat{M}
\hat{P} \hat{M} \mathit{\Delta}_{3}^{(3)}S_{2}^{(3)} -
\frac{3}{2}\,\mathit{\delta}_{11}^{(3)}\hat{P}\hat{M}\hat{P}_{x}
S_{2}^{(3)} - \mathit{\delta}_{11}^{(3)}\hat{P}_{xx} S_{2}^{(3)} -
\hat{P}\hat{M}_{x} \mathit{\Delta}_{4}^{(3)} S_{4}^{(3)} \\
& &   \hspace{-1mm} + \, \hat{P}_{x}\hat{M}
\mathit{\Delta}_{4}^{(3)} S_{4}^{(3)}  +
\frac{3}{2}\,\hat{P}\hat{M} \hat{P} \hat{M}
\mathit{\Delta}_{4}^{(3)} S_{4}^{(3)} - \frac{3}{2}\,
\mathit{\delta}_{11}^{(3)}\hat{P}_{x} \hat{M}\hat{P} S_{2}^{(3)},
\label{eqa100}
\end{eqnarray}
with
\begin{equation}
\hspace{-8mm}  \hat{M} = \big(\mathit{\delta}_{11}^{(3)} M +
2\,\mathit{\delta}_{11}^{(3)}S_{2}^{(3)}\big)
\big(\mathit{\Delta}_{4}^{(3)}\big)^{-1}, \ \ \ \hat{P} = \big(
\mathit{\Delta}_{4}^{(3)}P + 2\,\mathit{\Delta}_{3}^{(3)}
\big)/\mathit{\delta}_{11}^{(3)}. \label{eqa101}
\end{equation}

\textbf{Darboux transformation D}: The linear system~(\ref{eqa84})
is kept invariant by the gauge transformation
\begin{equation}
 \mathit{\tilde{\Phi}} = \big(\lambda^{2} \mathit{\Delta}^{(4)}-
\lambda \mathit{\Delta}^{(4)} S_{\bot}^{(4)}
-\mathit{\Delta}^{(4)} S_{\top}^{(4)}\big)\mathit{\Phi},
\label{eqa102}
\end{equation}
with
\begin{eqnarray}
 &  & \mathit{\Delta}^{(4)}=
\begin{pmatrix}
\mathit{\Delta}_{1}^{(4)} & \mathit{\Delta}_{2}^{(4)} \\
\mathit{\Delta}_{3}^{(4)} & \mathit{\Delta}_{4}^{(4)}
\end{pmatrix}, \ \ S_{\bot}^{(4)} = \begin{pmatrix}
0 & S_{2}^{(4)} \\
S_{3}^{(4)} & O_{2}
\end{pmatrix}, \ \
S_{\top}^{(4)} = \begin{pmatrix}
S_{1}^{(4)} & O_{1}^{T} \\
O_{1} & S_{4}^{(4)}
\end{pmatrix}
\label{eqb01}, \\
& & S_{\bot}^{(4)}H^{(4)}\Lambda^{(4)}\big(H^{(4)}\big)^{-1} +
S_{\top}^{(4)} =
H^{(4)}\big(\Lambda^{(4)}\big)^{2}\big(H^{(4)}\big)^{-1},
\label{eqb02}
\end{eqnarray}
where $
\Lambda^{(4)}={\rm{diag}}\big(\lambda_{1}^{(4)},\lambda_{2}^{(4)},\ldots,
\lambda_{N+1}^{(4)}\big) $, $
H^{(4)}=\big(h_{1}^{(4)},h_{2}^{(4)},\ldots, h_{N+1}^{(4)}\big) $
with $ h_{k}^{(4)} $  ($ 1 \leq k\leq N+1 $) as the column
solutions of System~(\ref{eqa84}) for different eigenvalues
$\lambda = \lambda_{k}^{(4)} $, $ \mathit{\Delta}_{2}^{(4)} =
O_{1}^{T} $, $ \mathit{\Delta}_{3}^{(4)} = O_{1} $, $
\mathit{\Delta}_{1}^{(4)} = \delta^{(4)}_{11} $, $
\mathit{\Delta}_{4}^{(4)} = \big( \delta^{(4)}_{ik}\big)_{2\leq
i,\, k \leq N+1} $ obey the following conditions:
\begin{eqnarray}
 & &  \hspace{-11mm}
\mathit{\delta}_{11,x}^{(4)} =
2\,\mathit{\delta}_{11}^{(4)}S_{2}^{(4)}S_{3}^{(4)} -
\mathit{\delta}_{11}^{(4)}S_{2}^{(4)}P -
\mathit{\delta}_{11}^{(4)}M S_{3}^{(4)}, \label{eqb03}
\\  & & \hspace{-11mm} \nonumber
\mathit{\delta}_{11,y}^{(4)}=
\mathit{\delta}_{11}^{(4)}\tilde{M}\tilde{P}S_{1}^{(4)} -
\mathit{\delta}_{11}^{(4)}S_{2}^{(4)}P M P -
\mathit{\delta}_{11}^{(4)}S_{2}^{(4)}P_{x} -
\mathit{\delta}_{11}^{(4)}S_{1}^{(4)}M P \\
& & \hspace{2mm} +\,
\tilde{M}_{x}\mathit{\Delta}_{4}^{(4)}S_{3}^{(4)} -
\tilde{M}\tilde{P}\tilde{M}\mathit{\Delta}_{4}^{(4)}S_{3}^{(4)},
\label{eqb04}
\\ & &  \hspace{-11mm} \nonumber
\mathit{\delta}_{11,t}^{(4)} = -
\frac{3}{2}\,\mathit{\delta}_{11}^{(4)}S_{2}^{(4)}P B -
\frac{3}{2}\,\mathit{\delta}_{11}^{(4)}S_{2}^{(4)}P_{x} M P
-\frac{3}{2}\,\mathit{\delta}_{11}^{(4)}S_{2}^{(4)}PMP_{x} -
\mathit{\delta}_{11}^{(4)}S_{2}^{(4)}P_{xx}   \\
& &  \nonumber \hspace{2mm} +\,
\mathit{\delta}_{11}^{(4)}\tilde{M}\tilde{P}_{x}S_{1}^{(4)} -
\frac{3}{2}\, \mathit{\delta}_{11}^{(4)}S_{1}^{(4)} B -
\mathit{\delta}_{11}^{(4)}S_{1}^{(4)}M P_{x}  + \frac{3}{2}
\tilde{M}_{x}\tilde{P}\tilde{M}\mathit{\Delta}_{4}^{(4)}S_{3}^{(4)} \\
& &  \nonumber \hspace{2mm} +\,
\frac{3}{2}\,\tilde{M}\tilde{P}\tilde{M}_{x}\mathit{\Delta}_{4}^{(4)}S_{3}^{(4)}
- \mathit{\delta}_{11}^{(4)}\tilde{M}_{x}\tilde{P}S_{1}^{(4)} -
\frac{3}{2}\,\tilde{B}\,\tilde{M}\mathit{\Delta}_{4}^{(4)}S_{3}^{(4)}
 +
\frac{3}{2}\,\mathit{\delta}_{11}^{(4)}\tilde{B} S_{1}^{(4)}
 \\
& &  \hspace{2mm}  +\, \mathit{\delta}_{11}^{(4)}S_{1}^{(4)}M_{x}
P - \tilde{M}_{xx}\mathit{\Delta}_{4}^{(4)}S_{3}^{(4)},
\label{eqb05}
\end{eqnarray}
\vspace{-11.5mm}
\begin{eqnarray}
 & &  \hspace{-25mm}
\mathit{\Delta}_{4,x}^{(4)} =
\mathit{\Delta}_{4}^{(4)}S_{3}^{(4)}M + \mathit{\Delta}_{4}^{(4)}P
S_{2}^{(4)}- 2\,\mathit{\Delta}_{4}^{(4)}S_{3}^{(4)}S_{2}^{(4)},
\label{eqb06}
\\  & & \hspace{-25mm} \nonumber
\mathit{\Delta}_{4,y}^{(4)}= \mathit{\Delta}_{4}^{(4)}S_{3}^{(4)}M
P M - \mathit{\Delta}_{4}^{(4)}S_{3}^{(4)}M_{x} +
\mathit{\Delta}_{4}^{(4)}S_{4}^{(4)}P M -
\tilde{P}\tilde{M}\mathit{\Delta}_{4}^{(4)}S_{4}^{(4)}  \\
& & \hspace{-12mm} +\,
\mathit{\delta}_{11}^{(4)}\tilde{P}_{x}S_{2}^{(4)} +
\mathit{\delta}_{11}^{(4)}\tilde{P}\tilde{M}\tilde{P}S_{2}^{(4)},
\label{eqb07}
\end{eqnarray}
\begin{eqnarray}
& &  \hspace{-8mm} \nonumber \mathit{\Delta}_{4,t}^{(4)} =
\frac{3}{2}\,\mathit{\Delta}_{4}^{(4)}S_{3}^{(4)}B M -
\frac{3}{2}\,\mathit{\Delta}_{4}^{(4)}S_{3}^{(4)}M P M_{x}
-\frac{3}{2}\,\mathit{\Delta}_{4}^{(4)}S_{3}^{(4)}M_{x}PM +
\mathit{\Delta}_{4}^{(4)}S_{3}^{(4)}M_{xx}   \\
& &  \nonumber \hspace{4.5mm} +\,
\mathit{\Delta}_{4}^{(4)}S_{4}^{(4)}P_{x}M -
\mathit{\Delta}_{4}^{(4)}S_{4}^{(4)}PM_{x}  + \frac{3}{2}\,
\mathit{\delta}_{11}^{(4)}\tilde{P}\tilde{M}\tilde{P}_{x}S_{2}^{(4)}
+
\frac{3}{2}\,\mathit{\Delta}_{4}^{(4)}S_{4}^{(4)}PMPM\\
& &  \nonumber \hspace{4.5mm} +
\frac{3}{2}\,\mathit{\delta}_{11}^{(4)}\tilde{P}\tilde{B}S_{2}^{(4)}
+ \frac{3}{2}\,
\mathit{\delta}_{11}^{(4)}\tilde{P}_{x}\tilde{M}\tilde{P}S_{2}^{(4)}
-\frac{3}{2}\,\tilde{P}\tilde{M}\tilde{P}\tilde{M}\mathit{\Delta}_{4}^{(4)}S_{4}^{(4)}
-  \tilde{P}_{x} \tilde{M}\mathit{\Delta}_{4}^{(4)}S_{4}^{(4)}
 \\
& &  \hspace{4.5mm}  +\,\tilde{P}
\tilde{M}_{x}\mathit{\Delta}_{4}^{(4)}S_{4}^{(4)} +
\mathit{\delta}_{11}^{(4)}\tilde{P}_{xx}S_{2}^{(4)}, \label{eqb08}
\end{eqnarray}
with
\begin{equation}
\hspace{-5mm} \tilde{M} = \big(\mathit{\delta}_{11}^{(4)}M -
2\,\mathit{\delta}_{11}^{(4)}S_{2}^{(4)}\big)\big(\mathit{\Delta}_{4}^{(4)}\big)^{-1},
\ \ \  \tilde{P} = \big( \mathit{\Delta}_{4}^{(4)}P -
2\,\mathit{\Delta}_{4}^{(4)}S_{3}^{(4)}\big)/\mathit{\delta}_{11}^{(4)}.
\label{eqb09}
\end{equation}

\vspace{7mm}

\noindent {\textbf{5. New solitary-wave solutions with symbolic
computation}}

From the previous results, we can gain a series of explicit
solutions for the mKP equation~(\ref{eqa01}) by the following
iterative procedure:

(1) For the initial potentials $ (U,V) $ and $ (M,P) $, solve the
linear systems~(\ref{eqa14}), (\ref{eqa22}), (\ref{eqa77})
and~(\ref{eqa84}) with different eigenvalues $ \lambda_{k}^{(i)} $
for column solutions   $ h_{k}^{(i)}  $   ($ 1 \leq k\leq N+1 $; $
1 \leq i\leq 4 $).

(2) Work out the matrices $ S^{(1)}$, $ S_{\bot}^{(2)}$, $
S_{\top}^{(2)}$,  $ S^{(3)}$, $ S_{\bot}^{(4)}$, $ S_{\top}^{(4)}$
and  $ \mathit{\Delta}^{(i)}$ ($ 1 \leq i\leq 4 $), and yield the
new potentials $ (\hat{U}, \hat{V}) $, $ (\tilde{U}, \tilde{V} )
$, $ (\hat{M}, \hat{P}) $ and $ (\tilde{M}, \tilde{P}) $ via
Transformations~(\ref{eqa37}), (\ref{eqa55}), (\ref{eqa101}) and
(\ref{eqb09}).

(3) Substitute $ (\hat{U}, \hat{V}) $, $ (\tilde{U}, \tilde{V} ) $
into Expression~(\ref{eqa06}) and $ (\hat{M}, \hat{P}) $, $
(\tilde{M}, \tilde{P}) $ into Expression~(\ref{eqa09}), and obtain
four families of solutions for the mKP equation~(\ref{eqa01}) of
the form
\renewcommand{\theequation}{5.\arabic{equation}}
\setcounter{equation}{0}
\begin{eqnarray}
& & q_{N}^{(1)} = -\frac{1}{2}\,\hat{U} \hat{V} =
-\frac{1}{2}\,\big(U -
2\,S_{2}^{(1)}\big)\big[2\,\big(\mathit{\Delta}_{4}^{(1)}\big)^{-1}\mathit{\Delta}_{3}^{(1)}
+ V \big],  \label{eqb51} \\
& & q_{N}^{(2)} = -\frac{1}{2}\, \tilde{U} \tilde{V} =
-\frac{1}{2}\,\big(U -
2\,S_{2}^{(2)}\big)\big(V + 2\,S_{3}^{(2)} \big), \label{eqb52} \\
& & q_{N}^{(3)} = -\frac{1}{2}\,\hat{M} \hat{P} = -\frac{1}{2}\,
\big(M + 2\,S_{2}^{(3)}\big) \big[
2\,\big(\mathit{\Delta}_{4}^{(3)}\big)^{-1}\mathit{\Delta}_{3}^{(3)}
+ P \big], \label{eqb53} \\
& & q_{N}^{(4)} = -\frac{1}{2}\,\tilde{M}\tilde{P} =
-\frac{1}{2}\, \big(M - 2\,S_{2}^{(4)}\big)\big( P -
2\,S_{3}^{(4)}\big). \label{eqb54}
\end{eqnarray}

Based on the above explained procedure, in what follows we shall
combine Darboux transformations A--D with Decompositions I and II
to construct explicit solutions of Eqn.~(\ref{eqa01}) by starting
with the trivial seed solutions of
Systems~(\ref{eqa07})--(\ref{eqa08}) and
(\ref{eqa10})--(\ref{eqa11}). For instance, we solve the linear
system~(\ref{eqa14}) with $ N=1 $, $ u_{1} = v_{1} =0 $, $ \lambda
= \lambda^{(1)}_{k} $ $ (k=1,2;  \lambda^{(1)}_{1} \neq
\lambda^{(1)}_{2}) $ and get the following:
\begin{subequations}
\begin{align}
&  h^{(1)}_{11} =\alpha^{(1)}_1 \text{exp} \big[\lambda^{(1)}_1 x
+ 2\,\big({\lambda^{(1)}_1}\big)^2 y +
4\,\big(\lambda^{(1)}_1\big)^3 t \big],  \label{eqb55a}\\
&  h^{(1)}_{21} =\alpha^{(1)}_2 \text{exp}\big[-\lambda^{(1)}_1 x
- 2\,\big({\lambda^{(1)}_1}\big)^2 y -
4\,\big(\lambda^{(1)}_1\big)^3 t \big], \label{eqb55b} \\
&  h^{(1)}_{12} =\alpha^{(1)}_3 \text{exp} \big[\lambda^{(1)}_2 x
+ 2\,\big({\lambda^{(1)}_2}\big)^2 y +
4\,\big(\lambda^{(1)}_2 \big)^3 t \big],  \label{eqb55c}\\
&  h^{(1)}_{22} =\alpha^{(1)}_4 \text{exp}\big[-\lambda^{(1)}_2 x
- 2\,\big({\lambda^{(1)}_2}\big)^2 y - 4\,\big(\lambda^{(1)}_2
\big)^3 t \big], \label{eqb55d}
\end{align}  \label{eqb55}
\end{subequations}
\hspace{-2mm}where $ \lambda^{(1)}_{1} $, $ \lambda^{(1)}_{2} $
and $ \alpha^{(1)}_{l} $ $ (l=1,2,3,4) $ are all nonzero
constants. From Eqn.~(\ref{eqa47}),  the element $ s^{(1)}_{12} $
of matrix $ S^{(1)} $ is expressed as
\begin{equation}
s^{(1)}_{12} = \frac{\big(\lambda^{(1)}_1-\lambda^{(1)}_2 \big)
h^{(1)}_{11}h^{(1)}_{12} }{ h^{(1)}_{12}h^{(1)}_{21}-
h^{(1)}_{11}h^{(1)}_{22} }. \label{eqb56}
\end{equation}
Proceedingly,  by performing symbolic manipulations on
Eqns.~(\ref{eqa35}), (\ref{eqa41}) and~(\ref{eqa42}) with
substitution of Expressions~(\ref{eqb55a})--(\ref{eqb55d}), the
function $ \delta^{(1)}_{21} $ is figured out as follows:
\begin{equation}
\delta^{(1)}_{21} = \frac{\delta^{(1)}_{22} \alpha^{(1)}_1
\alpha^{(1)}_2 \alpha^{(1)}_3 \alpha^{(1)}_4 \big(\lambda^{(1)}_1
- \lambda^{(1)}_2 \big)}{\alpha^{(1)}_1 \alpha^{(1)}_2
\lambda^{(1)}_2 \big(h^{(1)}_{12}\big)^{2}  - \alpha^{(1)}_3
\alpha^{(1)}_4 \lambda^{(1)}_1 \big(h^{(1)}_{11}\big)^{2} },
\label{eqb57}
\end{equation}
with $ \delta^{(1)}_{22} $ as an arbitrary nonzero constant. To
this point, the first family of solitary-wave solutions for
Eqn.~(\ref{eqa01}) is obtained as
\begin{equation}
\hspace{-72mm} q_{1}^{(1)} = \frac{2\,\alpha^{(1)}_1
\alpha^{(1)}_2 \alpha^{(1)}_3 \alpha^{(1)}_4\big(\lambda^{(1)}_1 -
\lambda^{(1)}_2 \big)^2\text{sech}^2 \xi^{(1)} }{\big(
\gamma^{(1)}_{1}  +
 \gamma^{(1)}_{3}  \tanh
\xi^{(1)} \big)\big( \gamma^{(1)}_{2}  +  \gamma^{(1)}_{4}  \tanh
\xi^{(1)} \big)}, \label{eqb58}
\end{equation}
\hspace{1mm} with
\begin{eqnarray}
\hspace{-32.5mm}
\begin{array}{l}
\vspace{1mm}
 \xi^{(1)} = \big(\lambda^{(1)}_1 - \lambda^{(1)}_2 \big)x + 2
\big[\big(\lambda^{(1)} _1\big)^{2}-\big(\lambda^{(1)}_2\big)^{2}
\big]y + 4
\big[\big(\lambda^{(1)}_1\big)^{3}-\big(\lambda^{(1)}_2\big)^{3}
\big] t,    \\
\vspace{1mm}
 \gamma^{(1)}_{1} = \alpha^{(1)}_1 \alpha^{(1)}_4-\alpha^{(1)}_2 \alpha^{(1)}_3, \ \
\gamma^{(1)}_{2} = \alpha^{(1)}_1 \alpha^{(1)}_4 \lambda^{(1)}_1 -
\alpha^{(1)}_2 \alpha^{(1)}_3 \lambda^{(1)}_2,  \\
 \gamma^{(1)}_{3} = \alpha^{(1)}_2 \alpha^{(1)}_3+\alpha^{(1)}_1 \alpha^{(1)}_4, \ \
\gamma^{(1)}_{4} =  \alpha^{(1)}_1 \alpha^{(1)}_4 \lambda^{(1)}_1
+ \alpha^{(1)}_2 \alpha^{(1)}_3 \lambda^{(1)}_2.
\end{array} \label{eqb59}
\end{eqnarray}
In a like manner, we can present other three families of
solitary-wave solutions for Eqn.~(\ref{eqa01}):
\begin{equation}
\hspace{-83mm} q_{1}^{(2)} =
\frac{2\,\gamma^{(2)}_{1}\big[\big(\lambda^{(2)}_1\big)^2 -
\big(\lambda^{(2)}_2 \big)^2 \big] \text{sech}\,2\,\xi^{(2)}}{
 \gamma^{(2)}_{2} \text{sech}\,2\,\xi^{(2)}  + \gamma^{(2)}_{3}
\tanh 2\,\xi^{(2)}- \gamma^{(2)}_{4}}, \label{eqb510}
\end{equation}
\hspace{1mm} with
\begin{eqnarray}
\hspace{-5mm}
\begin{array}{l}
\vspace{1mm}
 \xi^{(2)} = -\big[ \big( \lambda^{(2)}_1 \big)^{2} -\big( \lambda^{(2)}_2 \big)^{2} \big]x + 2
\big[\big(\lambda^{(2)} _1\big)^{4}-\big(\lambda^{(2)}_2\big)^{4}
\big]y -
4\big[\big(\lambda^{(2)}_1\big)^{6}-\big(\lambda^{(2)}_2\big)^{6}
\big] t,    \\
\vspace{1mm}
 \gamma^{(2)}_{1} = \alpha^{(2)}_1
\alpha^{(2)}_2 \alpha^{(2)}_3 \alpha^{(2)}_4
\big[\big(\lambda^{(2)}_1\big)^2 - \big(\lambda^{(2)}_2 \big)^2
\big], \ \ \gamma^{(2)}_{2} = \alpha^{(2)}_1 \alpha^{(2)}_2
\alpha^{(2)}_3 \alpha^{(2)}_4
\big[\big(\lambda^{(2)}_1\big)^2 + \big(\lambda^{(2)}_2 \big)^{2} \big],  \\
 \gamma^{(2)}_{3} = \big[\big(\alpha^{(2)}_2 \alpha^{(2)}_3 \big)^2
- \big(\alpha^{(2)}_1 \alpha^{(2)}_4 \big)^2\big]
\lambda^{(2)}_1\lambda^{(2)}_2, \ \ \gamma^{(2)}_{4} =
\big[\big(\alpha^{(2)}_2 \alpha^{(2)}_3 \big)^2 +
\big(\alpha^{(2)}_1 \alpha^{(2)}_4 \big)^2\big]
\lambda^{(2)}_1\lambda^{(2)}_2,
\end{array} \label{eqb511}
\end{eqnarray}

\begin{equation}
\hspace{-70mm} q_{1}^{(3)} = \frac{-2\,\alpha^{(3)}_1
\alpha^{(3)}_2 \alpha^{(3)}_3 \alpha^{(3)}_4\big(\lambda^{(3)}_1 -
\lambda^{(3)}_2 \big)^2\text{sech}^2 \xi^{(3)} }{\big(
\gamma^{(3)}_{1}  +
 \gamma^{(3)}_{3}  \tanh
\xi^{(3)} \big)\big( \gamma^{(3)}_{2}  +  \gamma^{(3)}_{4}  \tanh
\xi^{(3)} \big)}, \label{eqb512}
\end{equation}
\hspace{3mm}with
\begin{eqnarray}
\hspace{-28mm}
\begin{array}{l}
\vspace{1mm}
 \xi^{(3)} = -\big(\lambda^{(3)}_1 - \lambda^{(3)}_2 \big)x + 2
\big[\big(\lambda^{(3)} _1\big)^{2}-\big(\lambda^{(3)}_2\big)^{2}
\big]y - 4
\big[\big(\lambda^{(3)}_1\big)^{3}-\big(\lambda^{(3)}_2\big)^{3}
\big] t,    \\
\vspace{1mm}
 \gamma^{(3)}_{1} = \alpha^{(3)}_1 \alpha^{(3)}_4-\alpha^{(3)}_2 \alpha^{(3)}_3, \ \
\gamma^{(3)}_{2} = \alpha^{(3)}_1 \alpha^{(3)}_4 \lambda^{(3)}_1 -
\alpha^{(3)}_2 \alpha^{(3)}_3 \lambda^{(3)}_2,  \\
 \gamma^{(3)}_{3} = \alpha^{(3)}_2 \alpha^{(3)}_3+\alpha^{(3)}_1 \alpha^{(3)}_4, \ \
\gamma^{(3)}_{4} =  \alpha^{(3)}_1 \alpha^{(3)}_4 \lambda^{(3)}_1
+ \alpha^{(3)}_2 \alpha^{(3)}_3 \lambda^{(3)}_2 ,
\end{array} \label{eqb513}
\end{eqnarray}

\begin{equation}
\hspace{-83mm} q_{1}^{(4)} =
\frac{-2\,\gamma^{(4)}_{1}\big[\big(\lambda^{(4)}_1\big)^2 -
\big(\lambda^{(4)}_2 \big)^2 \big] \text{sech}\,2\,\xi^{(4)}}{
 \gamma^{(4)}_{2} \text{sech}\,2\,\xi^{(4)}  + \gamma^{(4)}_{3}
\tanh 2\,\xi^{(4)}- \gamma^{(4)}_{4} },  \label{eqb514}
\end{equation}
\hspace{1mm} with
\begin{eqnarray}
\hspace{-4mm}
\begin{array}{l}
\vspace{1mm}
 \xi^{(4)} = \big[ \big( \lambda^{(4)}_1 \big)^{2} -\big( \lambda^{(4)}_2 \big)^{2} \big]x + 2
\big[\big(\lambda^{(4)} _1\big)^{4}-\big(\lambda^{(4)}_2\big)^{4}
\big]y +
4\big[\big(\lambda^{(4)}_1\big)^{6}-\big(\lambda^{(4)}_2\big)^{6}
\big] t,    \\
\vspace{1mm}
 \gamma^{(4)}_{1} = \alpha^{(4)}_1
\alpha^{(4)}_2 \alpha^{(4)}_3 \alpha^{(4)}_4
\big[\big(\lambda^{(4)}_1\big)^2 - \big(\lambda^{(4)}_2 \big)^2
\big], \ \ \gamma^{(4)}_{2} = \alpha^{(4)}_1 \alpha^{(4)}_2
\alpha^{(4)}_3 \alpha^{(4)}_4
\big[\big(\lambda^{(4)}_1\big)^2 + \big(\lambda^{(4)}_2 \big)^{2} \big],  \\
 \gamma^{(4)}_{3} = \big[\big(\alpha^{(4)}_1 \alpha^{(4)}_4 \big)^2 -
\big(\alpha^{(4)}_2 \alpha^{(4)}_3 \big)^2\big]
\lambda^{(4)}_1\lambda^{(4)}_2, \ \ \gamma^{(4)}_{4} =
\big[\big(\alpha^{(4)}_2 \alpha^{(4)}_3 \big)^2 +
\big(\alpha^{(4)}_1 \alpha^{(4)}_4 \big)^2\big]
\lambda^{(4)}_1\lambda^{(4)}_2 .
\end{array}  \label{eqb515}
\end{eqnarray}

We recall that $ q_{1}^{(i)} $ ($ i =1,2,3,4 $) are four new
families of solitary-wave solutions for Eqn.~(\ref{eqa01}) since
they are different from the conventional solitary-wave solutions
in terms of a finite series of tanh and/or sech functions. Through
the qualitative analysis, it is easy to see that the functions $
q_{1}^{(i)} $ ($ i =1,2,3,4 $) have no singularity and exhibit
stable bell profiles for $ \big( \alpha^{(i)}_1, \alpha^{(i)}_2,
\alpha^{(i)}_3, \alpha^{(i)}_4, \lambda^{(i)}_1, \lambda^{(i)}_2
\big)  \in \mathcal{A}^{(i)} \cup \mathcal{ B }^{(i)}  $, where $
\mathcal{A}^{(i)} $ and  $ \mathcal{ B }^{(i)} $ are defined by
\begin{eqnarray}
& & \hspace{-5mm} \nonumber  \mathcal{A}^{(1)}= \big\{  \big(
\alpha^{(1)}_1, \alpha^{(1)}_2, \alpha^{(1)}_3, \alpha^{(1)}_4,
\lambda^{(1)}_1, \lambda^{(1)}_2 \big) \big|\alpha^{(1)}_2
\alpha^{(1)}_3 < 0, \alpha^{(1)}_1 \alpha^{(1)}_4 > 0,
\lambda^{(1)}_1 \lambda^{(1)}_2>0, \lambda^{(1)}_1 \neq
\lambda^{(1)}_2 \big\}, \\
& & \hspace{-5mm} \nonumber  \mathcal{B}^{(1)}= \big\{  \big(
\alpha^{(1)}_1, \alpha^{(1)}_2, \alpha^{(1)}_3, \alpha^{(1)}_4,
\lambda^{(1)}_1, \lambda^{(1)}_2 \big) \big|\alpha^{(1)}_2
\alpha^{(1)}_3 > 0, \alpha^{(1)}_1 \alpha^{(1)}_4 < 0,
\lambda^{(1)}_1 \lambda^{(1)}_2>0, \lambda^{(1)}_1 \neq
\lambda^{(1)}_2 \big\},  \\
& & \hspace{-5mm} \nonumber  \mathcal{A}^{(2)}= \big\{  \big(
\alpha^{(2)}_1, \alpha^{(2)}_2, \alpha^{(2)}_3, \alpha^{(2)}_4,
\lambda^{(2)}_1, \lambda^{(2)}_2 \big) \big| \alpha^{(2)}_1
\alpha^{(2)}_2 \alpha^{(2)}_3 \alpha^{(2)}_4 < 0, \lambda^{(2)}_1
\lambda^{(2)}_2>0, \lambda^{(2)}_1 \neq \lambda^{(2)}_2 \big\}, \\
& & \hspace{-5mm} \nonumber  \mathcal{B}^{(2)}= \big\{  \big(
\alpha^{(2)}_1, \alpha^{(2)}_2, \alpha^{(2)}_3, \alpha^{(2)}_4,
\lambda^{(2)}_1, \lambda^{(2)}_2 \big) \big| \alpha^{(2)}_1
\alpha^{(2)}_2 \alpha^{(2)}_3 \alpha^{(2)}_4 > 0, \lambda^{(2)}_1
\lambda^{(2)}_2 < 0, \lambda^{(2)}_1 \neq \lambda^{(2)}_2 \big\},
\\
& & \hspace{-5mm} \nonumber  \mathcal{A}^{(3)}= \big\{  \big(
\alpha^{(3)}_1, \alpha^{(3)}_2, \alpha^{(3)}_3, \alpha^{(3)}_4,
\lambda^{(3)}_1, \lambda^{(3)}_2 \big) \big|\alpha^{(3)}_2
\alpha^{(3)}_3 < 0, \alpha^{(3)}_1 \alpha^{(3)}_4 > 0,
\lambda^{(3)}_1 \lambda^{(3)}_2>0, \lambda^{(3)}_1 \neq
\lambda^{(3)}_2 \big\}, \\
& & \hspace{-5mm} \nonumber  \mathcal{B}^{(3)}= \big\{  \big(
\alpha^{(3)}_1, \alpha^{(3)}_2, \alpha^{(3)}_3, \alpha^{(3)}_4,
\lambda^{(3)}_1, \lambda^{(3)}_2 \big) \big|\alpha^{(3)}_2
\alpha^{(3)}_3 > 0, \alpha^{(3)}_1 \alpha^{(3)}_4 < 0,
\lambda^{(3)}_1 \lambda^{(3)}_2>0, \lambda^{(3)}_1 \neq
\lambda^{(3)}_2 \big\}, \\
& & \hspace{-5mm} \nonumber  \mathcal{A}^{(4)}= \big\{  \big(
\alpha^{(4)}_1, \alpha^{(4)}_2, \alpha^{(4)}_3, \alpha^{(4)}_4,
\lambda^{(4)}_1, \lambda^{(4)}_2 \big) \big| \alpha^{(4)}_1
\alpha^{(4)}_2 \alpha^{(4)}_3 \alpha^{(4)}_4 < 0, \lambda^{(4)}_1
\lambda^{(4)}_2>0, \lambda^{(4)}_1 \neq \lambda^{(4)}_2 \big\}, \\
& & \hspace{-5mm} \nonumber  \mathcal{B}^{(4)}= \big\{  \big(
\alpha^{(4)}_1, \alpha^{(4)}_2, \alpha^{(4)}_3, \alpha^{(4)}_4,
\lambda^{(4)}_1, \lambda^{(4)}_2 \big) \big| \alpha^{(4)}_1
\alpha^{(4)}_2 \alpha^{(4)}_3 \alpha^{(4)}_4 > 0, \lambda^{(4)}_1
\lambda^{(4)}_2 < 0, \lambda^{(4)}_1 \neq \lambda^{(4)}_2 \big\}.
\end{eqnarray}

\vspace{4mm}

\noindent {\textbf{6. Conclusions and discussions}

In this paper, we have shown that the mKP equation~(\ref{eqa01})
can be reduced to  the first two nontrivial nonlinear systems in
the $ 2N $-coupled CLL and KN hierarchies, i.e.,
Systems~(\ref{eqa07})--(\ref{eqa08}) and
(\ref{eqa10})--(\ref{eqa11}), by imposing the potential
constraints~(\ref{eqa06}) and~(\ref{eqa09}) on
Systems~(\ref{eqa02})--(\ref{eqa03})
and~(\ref{eqa04})--(\ref{eqa05}), respectively. Furthermore, it
has been found that the $ 2N $-coupled CLL and high-order CLL
systems possess two different Lax representations~(\ref{eqa14})
and~(\ref{eqa22}), and the $ 2N $-coupled KN and high-order KN
systems also admit  two different Lax
representations~(\ref{eqa77}) and~(\ref{eqa84}). For these four
Lax representations, we have constructed the corresponding Darboux
transformations by which abundant explicit solutions of
Eqn.~(\ref{eqa01}) can be obtained in a recursive manner. Through
one-time iteration of those Darboux transformations, four new
families of solitary-wave solutions have been presented and the
relevant stability has been analyzed. Finally, we would like to
discuss the following issues:

\begin{enumerate}

\item

As far as we know, there are usually two ways of finding the
integrable decompositions for a (2+1)-dimensional integrable NLEE:
the first is to choose a proper (1+1)-dimensional soliton
hierarchy and relate its first two nontrivial members to the
desired (2+1)-dimensional equation~\cite{a15,a21a,a13}; the second
is to directly nonlinearize one or two Lax pairs of the
(2+1)-dimensional integrable equation into two (1+1)-dimensional
nonlinear systems~\cite{a09a,a10,a11,a12,a21a}. Obviously, the
proposal of Decompositions I and II in Section 2 is based on the
second decomposition method. In addition, it is noted that the
potential constraint~(\ref{eqa06}) or~(\ref{eqa09}), which
originates from the Bargmann symmetry constraint~\cite{a30}, might
also be applicable to some other (2+1)-dimensional integrable
NLEEs with the availability of two symmetry Lax pairs like
Systems~(\ref{eqa02})--(\ref{eqa03})
and~(\ref{eqa04})--(\ref{eqa05}). Special attention should be paid
to that Systems~(\ref{eqa02}) and~(\ref{eqa03}) can be transformed
to each other with  $ \left[ q(x,y,t), u(x,y,t) \right]
\longleftrightarrow \left[q(-x,-y,-t), v(-x,-y,-t)\right] $, so do
Systems~(\ref{eqa04}) and~(\ref{eqa05}) with  $ \left[ q(x,y,t),
m(x,y,t) \right] $ $ \longleftrightarrow \left[q(-x,-y,-t),
p(-x,-y,-t)\right] $.

\item

From the derivation of Darboux transformations A--D, we infer that
for the following two general linear spectral problems
\renewcommand{\theequation}{6.\arabic{equation}}
\setcounter{equation}{0}
\begin{eqnarray}
& & \hspace{-2mm} \mathit{\Psi}_{x}= \big[\lambda\,Q^{(1)}_{0} +
Q^{(1)}_{1} \big] \mathit{\Psi},  \label{eqb601}
\\ & &  \hspace{-2mm}
\mathit{\Phi}_{y} =  \big[ \lambda^{2}\,Q^{(2)}_{0} +
\lambda\,Q^{(2)}_{1}+ Q^{(2)}_{2}  \big]\!\mathit{\Phi},
\label{eqb602}
\end{eqnarray}
where $ \lambda $ is the eigenvalue parameter, $ \mathit{\Psi}\!
=\! (\psi_{1}, \psi_{2}, \ldots, \psi_{N})^{T} $, $
\mathit{\Phi}\! =\! (\phi_{1}, \phi_{2}, \ldots, \phi_{N})^{T} $
are the vector eigenfunctions,  $ Q^{(1)}_{i} $ and  $ Q^{(2)}_{k}
$ ($ i=0,1; k=0,1,2 $) are all the $ N \times N $ matrices, the
corresponding Darboux transformations can be respectively taken as
\begin{eqnarray}
& &  \mathit{\hat{\Psi}} = (\lambda\,\mathit{\Delta}_{1}
-\mathit{\Delta}_{1} S )\mathit{\Psi},  \label{eqb603} \\
& & \mathit{\hat{\Phi}} = (\lambda^{2} \mathit{\Delta}_{2}-
\lambda\, \mathit{\Delta}_{2} S_{\bot} -\mathit{\Delta}_{2}
S_{\top} )\mathit{\Phi}, \label{eqb604}
\end{eqnarray}
where $ \mathit{\Delta}_{1} $, $ \mathit{\Delta}_{2} $, $ S $, $
S_{\bot} $ and $ S_{\top} $ are five $ N \times N $ undetermined
matrices, $ \mathit{\hat{\Psi}} $ and $ \mathit{\hat{\Phi}} $
respectively satisfy Eqns.~(\ref{eqb601}) and~(\ref{eqb602}) with
$ Q^{(1)}_{i} $ and $ Q^{(2)}_{k} $ replaced by $
\hat{Q}^{(1)}_{i} $ and $ \hat{Q}^{(2)}_{k} $ ($ i=0,1; k=0,1,2
$). By utilizing Ans\"{a}tzs~(\ref{eqb603}) and~(\ref{eqb604}),
one can attempt to construct the Darboux transformations for many
hierarchies of soliton equations.

\item

It is emphasized that the iterative algorithm of the Darboux
transformation can be easily achieved on the computerized symbolic
computation systems such as \textit{Mathematica} and
\textit{Maple}. Accordingly, if we choose $ N>1 $ and make the
successive iteration of Darboux transformations A--D, many more
complicated explicit solutions of Eqn.~(\ref{eqa01}) will be
unearthed, which might be different from those previously
obtained.
\end{enumerate}

$\,$

\noindent {\bf Acknowledgements}

We would like to thank Profs.\ Y.\ T.\ Gao, F.\ W.\ Sun, G.\ M.\
Wei and Y.\ P.\ Liu for their valuable comments. We are also
grateful to the helpful discussion with Mr.\ W.\ Hu. This work has
been supported by the Key Project of Chinese Ministry of Education
(No.\ 106033), by the Specialized Research Fund for the Doctoral
Program of Higher Education (No.\ 20060006024), Chinese Ministry
of Education, and by the National Natural Science Foundation of
China under Grant No.\ 60372095.

\end{document}